\documentclass[fleqn,usenatbib]{mnras}

\usepackage{newtxtext,newtxmath}

\usepackage[T1]{fontenc}
\usepackage{ae,aecompl}

\usepackage{graphicx}	
\usepackage{amsmath}	
\usepackage{amssymb}	
\usepackage{lineno}

\usepackage{eso-pic}

\AddToShipoutPictureBG*{%
  \AtPageUpperLeft{%
    \hspace{0.75\paperwidth}%
    \raisebox{-3.5\baselineskip}{%
      \makebox[0pt][l]{\textnormal{DES 2018-0354}}
}}}%

\AddToShipoutPictureBG*{%
  \AtPageUpperLeft{%
    \hspace{0.75\paperwidth}%
    \raisebox{-4.5\baselineskip}{%
      \makebox[0pt][l]{\textnormal{FERMILAB-PUB-19-627-AE}}
}}}%

\title[STRIDES 2017]{The STRong lensing Insights into the Dark Energy Survey (STRIDES) 2017/2018 follow-up campaign: Discovery of 10 lensed quasars and 10 quasar pairs}

\author[C. A. Lemon et al.]{C. Lemon$^{1, 2, 3}$\thanks{E-mail: clemon@ast.cam.ac.uk}, M. W. Auger$^{1,2}$, R. McMahon$^{1,2}$, T. Anguita$^{4,5}$, Y. Apostolovski$^{4}$, \newauthor G. C.-F. Chen$^{6}$, C. D. Fassnacht$^{6}$, A.~Melo$^{7}$, V. Motta$^{7}$, A. Shajib$^{8}$, T. Treu$^{8}$, \newauthor A. Agnello$^{9}$, E. Buckley-Geer$^{10}$, P. L. Schechter$^{11}$, S. Birrer$^{8,12}$, T. Collett$^{13}$, \newauthor F. Courbin$^{3}$, C. E. Rusu$^{14, 15}$, T.~M.~C.~Abbott$^{16}$, S.~Allam$^{10}$, J.~Annis$^{10}$, S.~Avila$^{17}$, \newauthor E.~Bertin$^{18,19}$, D.~Brooks$^{20}$, D.~L.~Burke$^{12,21}$, A.~Carnero~Rosell$^{22,23}$, M.~Carrasco~Kind$^{24,25}$, \newauthor J.~Carretero$^{26}$, M.~Costanzi$^{27,28}$, L.~N.~da Costa$^{23,29}$,   J.~De~Vicente$^{22}$, S.~Desai$^{30}$,\newauthor T.~F.~Eifler$^{31,32}$, B.~Flaugher$^{10}$, J.~Frieman$^{10,33}$, J.~Garc\'ia-Bellido$^{17}$, E.~Gaztanaga$^{34,35}$, \newauthor D.~W.~Gerdes$^{36,37}$, D.~Gruen$^{12,21,38}$, R.~A.~Gruendl$^{24,25}$, J.~Gschwend$^{23,29}$, G.~Gutierrez$^{10}$, \newauthor K.~Honscheid$^{39,40}$, D.~J.~James$^{41}$, A.~Kim$^{42}$, E.~Krause$^{31}$, K.~Kuehn$^{43,44}$, N.~Kuropatkin$^{10}$, \newauthor O.~Lahav$^{20}$, M.~Lima$^{23,45}$, H.~Lin$^{10}$, M.~A.~G.~Maia$^{23,29}$, M.~March$^{46}$, J.~L.~Marshall$^{47}$, \newauthor F.~Menanteau$^{24,25}$,  R.~Miquel$^{26,48}$, A.~Palmese$^{10,33}$, F.~Paz-Chinch\'{o}n$^{24,25}$, A.~A.~Plazas$^{49}$, \newauthor A.~Roodman$^{12,21}$, E.~Sanchez$^{22}$, M.~Schubnell$^{37}$, S.~Serrano$^{34,35}$,  M.~Smith$^{50}$, \newauthor M.~Soares-Santos$^{51}$, E.~Suchyta$^{52}$, G.~Tarle$^{37}$, A.~R.~Walker$^{16}$
\\Affiliations are listed at the end of the paper
}

\date{Accepted XXX. Received YYY; in original form ZZZ}

\pubyear{2015}

\begin{document}
\label{firstpage}
\pagerange{\pageref{firstpage}--\pageref{lastpage}}
\maketitle
\newcommand{\MINUS}{\kern 0.15em --\kern 0.15em }

\begin{abstract}
We report the results of the STRong lensing Insights from the Dark Energy Survey (STRIDES) follow-up campaign of the late 2017/early 2018 season. We obtained spectra of 65 lensed quasar candidates either with EFOSC2 on the NTT or ESI on Keck, which confirm 10 new gravitationally lensed quasars and 10 quasar pairs with similar spectra, but which do not show a lensing galaxy in the DES images. Eight lensed quasars are doubly imaged with source redshifts between 0.99 and 2.90, one is triply imaged by a group and with all images detected by \textit{Gaia} (DESJ0345-2545, $z=$1.68), and one lens is quadruply imaged (quad: DESJ0053-2012, $z=$3.8). Singular isothermal ellipsoid models for the doubles, based on high-resolution imaging from SAMI on SOAR or NIRC2 on Keck, give total magnifications between 3.2 and 5.6, and Einstein radii ranging from 0.49 to 1.97 arcseconds. After spectroscopic follow-up, we extract multi-epoch \textit{grizY} photometry of confirmed lensed quasars and contaminant quasar+star pairs from the first 4 years of DES data using a parametric multi-band modelling routine and compare variability in each system's components. By measuring the reduced ${\chi}^2$ associated with fitting all epochs to the same magnitude, we find a simple cut on the less variable component that retains all confirmed lensed quasars, while removing 94 per cent of spectroscopically confirmed contaminant systems with stellar components. Based on our spectroscopic follow-up, this variability information can improve selection of lensed quasars and quasar pairs from 34-45 per cent to 51-70 per cent, with the majority of remaining contaminants being compact star-forming galaxies. Using mock lensed quasar lightcurves we demonstrate that selection based only on variability will over-represent the quad fraction by 10 per cent over a complete DES magnitude-limited sample (excluding microlensing differences), explained by the magnification bias and hence lower luminosity (more variable) sources in quads.
\end{abstract}




\section{Introduction}
The Dark Energy Survey (DES) has imaged over 5000 square degrees of extragalactic sky in the Southern Hemisphere in five optical to near-infrared bands: \textit{grizY}. Its depth and wide area provide the opportunity to discover rare objects \citep{desdr1}. Gravitationally lensed quasars are an example of such objects, with only $\sim$200 known to date. Only a subset of this sample can be applied to certain science cases: time-delay cosmography requires lenses with long time delays \citep[e.g.][]{saha2006}, well-separated images, and bright lensing galaxies \citep[e.g.][]{treu2002, suyu2014}; many microlensing studies ideally require pairs of images with short time delays \citep{bate2008, blackburne2011} or broad absorption line features \citep{sluse2015, hutsemekers2015, hutsemekers2019}. Furthermore, to exploit the unique capabilities of observatories in the Northern and Southern hemispheres, it is important to discover lenses across the sky. The known number of lensed quasars in the North celestial hemisphere is more than double that in the South: 143 to 60\footnote{https://www.ast.cam.ac.uk/ioa/research/lensedquasars/}, demonstrating the opportunity to mine lensed quasars from DES.

The STRong lensing Insights into the Dark Energy Survey
(STRIDES\footnote{STRIDES is a Dark Energy Survey Broad External Collaboration; PI: Tommaso Treu; http://strides.astro.ucla.edu}) was set up to find gravitationally lensed quasars in the DES footprint \citep{strides1} for use as cosmological probes, with particular efforts on measuring time delays via the COSMOGRAIL collaboration \citep[e.g.][]{courbin2018}. A sample of 40 well-studied time-delay lenses is expected to yield a 1 per cent precision measurement of the Hubble constant \citep{treu2016,shajib2018}. Since there is not yet a complementary spectroscopic survey of the DES footprint, lens searches must be based on purely photometric data. Given also the lack of a $u-$band in the DES wide field---a common component for selection of quasars below redshift $\sim$2.7--- efficient selection techniques must rely on other resources. Previous STRIDES searches have used DES photometry and mid-infrared colours from the \textit{Wide-field Infrared Survey Explorer} (\textit{WISE}) \citep{wright10}, coupled with: catalogue-based or pixel-based machine learning \citep{agnello2015}, component-fitting to pixels \citep{anguita18}, and \textit{Gaia} cross-matches to quasar candidates \citep[Ostrovski et al. in prep.,][]{agnello2018}. These searches have yielded $\sim$20 new gravitationally lensed quasars, including 4 quadruply imaged systems (hereafter quads), with one system already delivering time-delay cosmography inference \citep{shajib2019b}.

In this paper we present results from the STRIDES 2017-2018 follow-up campaign, pre-selecting candidates from the first three years of DES imaging data. In Section \ref{selection} we describe the lens selection techniques. Section \ref{results} details the results of the spectroscopic and imaging follow-up of individual candidates, with discussion of several systems. After spectroscopic follow-up, we implement a parametric modelling procedure on DES single-epoch images in Section \ref{variability} to derive variability properties of lensed quasars and contaminants, discussing the possibility of implementing this to refine selection and improve lens search efficiency. We summarise the paper in Section \ref{conclusion}.

\section{Lens Selection} \label{selection}
Given the variety of colours and morphologies of lensed quasars, individual selection techniques can be less effective in parts of this parameter space: for example, objects with bright lensing galaxies or quasars lensed by groups and clusters. It is therefore important to have several selection techniques each based on a different aspect of the DES data, or each including different external datasets. Our selection methods all rely on \textit{WISE} photometry, in particular the 3.4 and 4.6 micron bands (hereafter \textit{W1} and \textit{W2}). Lensed quasars are then selected via cross-matches to \textit{Gaia} data release 1 (DR1) (Section \ref{gaiaxm}), component fitting of the DES pixels (Section \ref{componentfitting}), or catalogued variability information from DES (Section \ref{variabilityselection}). The techniques and selection procedures are described in this section, with their spectroscopic follow-up outcomes given in Section \ref{results}.

\subsection{\textit{Gaia} Cross-matches} \label{gaiaxm}
\textit{Gaia}'s high-resolution (0.1 arcsecond FWHM), full-sky targeting of optically bright point sources makes it an ideal probe for resolving the multiple images of lensed quasars when ground-based data might blend them and catalogue a single object \citep{gaiadr12016, gaiadr120162}. \textit{Gaia} DR1 was released on 2016 September 14, providing a single broad-band optical magnitude and high-precision astrometry for 1,142,679,769 sources \citep{fabricius2016, lindegren2016}. We note that only \textit{Gaia} DR1 was used for these selections due to the release date, however \textit{Gaia} DR2 is more complete for detecting close-separation point sources \citep[e.g.,][]{arenou2018}.

\subsubsection{Method 1} \label{gaia1}
This selection method follows that of \citet{lemon2018}, relying on multiple \textit{Gaia} detections near \textit{WISE}-selected candidate quasars, or single \textit{Gaia} detections corresponding to extended DES objects. The latter technique was developed on account of \textit{Gaia} DR1 often cataloguing only single components of lensed quasars \citep{ducourant2018a}. The input catalogue for both searches is AllWISE detections with \textit{W1}$-$\textit{W2}$>$0.5, \textit{W1}$<$15.5, and catalogued uncertainties in \textit{W1} and \textit{W2}. The multiple-\textit{Gaia}-detection search required at least 2 \textit{Gaia} detections within 4 arcseconds of the \textit{WISE} source, and  within 5 arcseconds of each other. The single-\textit{Gaia}-detection search required one \textit{Gaia} detection within 4 arcseconds of an extended DES object ($MAG\_PSF\_I-MAG\_AUTO\_I>$ 0.2, $MAG\_AUTO\_I<$ 20.5). A stellar-density cut of < 50,000 \textit{Gaia} detections per square degree was applied to both techniques to remove star clusters, resulting in 5996 and 43,128 candidates, respectively. We note that the second data release of \textit{Gaia} has complete detection of bright images of doubly lensed quasars, increasing the efficiency of such cross-matched lens searches \citep{lemon2019}. Candidates were visually inspected and graded according to Section \ref{finalselection}.

\subsubsection{Method 2}
This method is described by \citet{agn17} and \citet{agnspi19}, using pre-selection from \textit{WISE} and astrometry from \textit{Gaia} in two ways. The first is based on \textit{WISE} preselection and deblending into \textit{multiplets}, either in the \textit{Gaia}-DR1 or DES catalogues. The second relies on  astrometric offsets of objects across \textit{Gaia} and 2MASS.

In the first search \citep[following][]{agn17}, objects with\footnote{Here, $WX=$\texttt{wXmpro} and $\delta WX=$\texttt{wXsigmpro} in the Vega system.}
\begin{equation}
    W1-W2>0.45+\sqrt{\delta W1^{2}+\delta W2^{2}}
\end{equation}
were matched to \textit{Gaia}-DR1, retaining only \textit{WISE} sources that corresponded to at least two \textit{Gaia} \texttt{source} entries within a $6^{\prime\prime}$ matching radius. After visual inspection, this resulted in the candidates and lenses presented by \citet[][]{agnello2018}, where numbers of objects, multiplets and candidates at each stage are reported, and whose confirmation spectra are shown in this paper. A subset of the initial objects did not correspond to \textit{Gaia} multiplets, but were deblended into multiple sources in the DES catalogue. After additional cuts in DES-\textit{WISE} colours and visual inspection, some were included in the multiplet sample (for further details we refer to \citet{agnello2018}, where they are presented).

The second search \citep[later described by][]{agnspi19} used the same \textit{WISE} preselection as above, but relied on the relative astrometry of the same sources in \textit{Gaia}-DR1 and 2MASS to isolate quasars with nearby companions that were not deblended into \textit{Gaia} multiplets. The main difference between that search and the implementation described by \citet{agnspi19} is that the latter used looser cuts in \textit{WISE} colours to include candidates with a significant lens-galaxy contribution to the photometry. This search re-discovered candidates that were selected as DES multiplets (among \textit{Gaia} singlets) in the first search.

\subsection{Component Fitting} \label{componentfitting}
This selection technique follows that of \citet{schechter17} and \citet{anguita18}. The initial list is taken from ALLWISE detections with \textit{W1}$<$14.5 and \textit{W1}$-$\textit{W2}$>$0.7. For each candidate, the best single-epoch image in each filter is fit as multiple quasi-Gaussians. To reject galaxies from the sample, quasi-Gaussians with widths larger than the local pipeline point spread function (PSF) are removed. For remaining systems with multiple bright components, two or three independent quasi-Gaussians are fit, and if at least two components are consistent with the local PSF, the system is retained. A colour selection, taking into account the possibility of reddening by a lensing galaxy, removes systems with very different optical colours, and a final selection is made by assigning Gaussian Mixture Model quasar probabilities from the \textit{griz} colours following \citet{ostrovski2017}. This search is complementary to the \textit{Gaia} searches as it is limited by the deeper DES depth. 

\subsection{Variability} \label{variabilityselection}
A number of physical processes have been proposed to explain quasar variability (see \citet{schmidt10} for examples and references). Irrespective of the physical process, we can use this intrinsic variability as a search technique. The technique was first proposed by \citet{schmidt10} as a way to detect quasars in wide-area sky surveys with no $u-$band imaging. They showed that with a survey such as the SDSS II Stripe 82, with $\sim$ 60 epochs of imaging data over 5 years, a variability selection could achieve a completeness and purity of 92 per cent, despite contaminants outnumbering true quasars by a factor of 30 in the input selection. Even with a much sparser sampling of 6 epochs over 3 years, such as with Pan-STARRS1 \citep[PS1,][]{chambers2016}, the variability method gave a 30 per cent pure and 75 per cent complete quasar candidate sample from a $griz$-selected input catalogue, which included the stellar locus, and a 92 per cent pure and 44 per cent complete quasar sample after removing the stellar locus. The method has also been successfully used to select a sample of high-redshift quasars in the SDSS III BOSS survey \citep{palanque11}. As the first three years of DES data have a similar number of epochs as PS1 \citep{diehl2016}, we should expect similar results for isolated quasars. We should note, however, that the behavior for lensed quasars may not be exactly the same as for isolated quasars due to microlensing.

For our lensed quasar search we follow the method described by \citet{schmidt10} and \citet{palanque11}.  In order to create an input sample on which to apply the variability algorithm we begin with the \textit{WISE} catalogue and select objects with \textit{W2} $< 14.45$ and \textit{W1} $-$ \textit{W2} $> 0.5$ on the Vega system. The selection yields 335,345 objects. For each object that is selected from the \textit{WISE} catalogue we then match it to an object in the DES Year 3 catalogue using a matching radius of 4 arcseconds. As the \textit{WISE} resolution is significantly worse than that of DES, there could be more than one DES object matched to the single \textit{WISE} detection. For each matched object we extract all the single-epoch detections of that object in $griz$ and use them to compute the variability. For each pair of observations $i$ and $j$ in a given band we define a magnitude difference $\Delta m_{i,j} = m_i - m_j$, where $m_i$ and $m_j$ are the magnitudes of the individual detections, and a time difference $\Delta t_{i,j} = t_i - t_j$, where $t_i$ and $t_j$ are the epochs of the observations. We assume a power law increase of the variability as a function of the time difference:

\begin{equation}
V_{\text{mod}}(\Delta t_{i,j}|A,\gamma) = A(\Delta t_{i,j}/1 \textrm{year})^{\gamma}
\label{eq:powerlaw}
\end{equation}
where $A$ is the variability amplitude, and $\gamma$ is the power-law index. We fit this model to the set of data $(\Delta m_{i,j},\Delta t_{i,j})$ by maximising $\mathcal{L} = \prod_{i,j} L_{i,j}$, where $L$ is defined as

\begin{equation}
    L_{i,j} = \frac{1}{\sqrt{2\pi V_{obs,i,j}^2}}\exp\left(-\frac{\Delta m_{i,j}^2}{2 V_{obs,i,j}^2}\right)
	\label{eq:likelihood}
\end{equation}
$V_{obs,i,j}^2$ is the observed variability and is equal to $V_{mod}(\Delta t_{i,j}|A,\gamma)^2 + (\sigma_i^2 + \sigma_j^2)$. The quantities $\sigma_i$ and $\sigma_j$ are the measured uncertainties on 
$m_i$ and $m_j$.
We require that an object is detected in at least 4 epochs in an individual band and then fit for $A$ and $\gamma$ for that band. We do this for all four bands. For objects that have at least 4 epochs and a good fit in at least 3 of the available bands we then follow the method of \citet{palanque11} and we fit for a single value of $\gamma$ in all bands and different values of $A$ ($A_g$, $A_r$, $A_i$, $A_z$).

For each object we require that it passes the following quasar selection cuts in three out of the four bands. 
\begin{multline}
\gamma > 0.5\log(A) + 0.50 \ , \ \gamma > -2\log(A) - 2.25 \ , \ \gamma > 0.055
\label{eq:qsocuts}
\end{multline}
This process results in 16,026 objects. These objects are then analyzed using the image splitting technique described in Section \ref{componentfitting}. We also looked explicitly at systems where there were two or more DES objects matched to the \textit{WISE} object and retained those where at least two of the objects passed the selection criteria of equation~\ref{eq:qsocuts} in at least three of the bands.

\begin{figure}
	\includegraphics[trim={0cm 1.5cm 1cm 2.5cm},clip,width=\columnwidth]{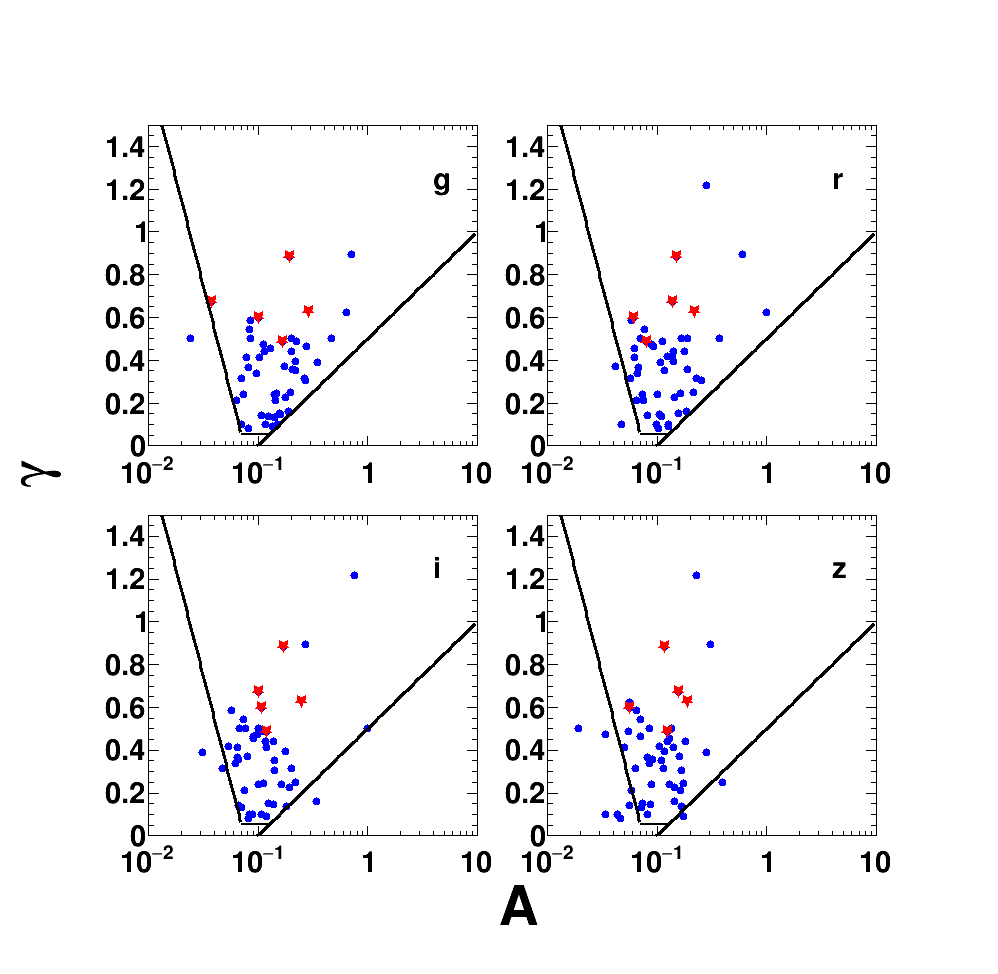}
    \caption{Distribution of variability amplitude, $A$, versus power-law index, $\gamma$, for the 56 variability-selected candidates. Upper left $g$-band; upper right $r$-band; lower left $i$-band; lower right $z$-band. Known lensed quasars are shown in red.}
    \label{fig:a_gamma}
\end{figure}

In total, we selected 56 new objects for spectroscopic and imaging follow-up. These objects were visually inspected. The method also recovered five known lensed quasars which were previously discovered by other STRIDES techniques. These five were not included in the 56. The 56 objects are shown in Fig.~\ref{fig:a_gamma} where we show the values of $A_g$, $A_r$, $A_i$, $A_z$ versus $\gamma$ for each object.

\subsection{Final Candidate Selection} \label{finalselection}
After visual inspection of each method's final candidate list by one author (\textit{Gaia} 1: CL, \textit{Gaia} 2: AA, Component Fitting: PS, Variability: EBG), any possible lensed quasars were given a ranking from 0-3. 3 was reserved for targets that must be followed up based on all photometric evidence suggesting it is a lens, 2 for probable lens candidates, 1 for possible lens candidates, and 0 for unlikely lens candidates.

While we still have many unobserved targets, all Rank 3 and most Rank 2 candidates were followed up either spectroscopically, or with high-resolution imaging, or both. This follow-up is detailed in Section \ref{results}.

\section{Results} \label{results}
\subsection{Spectroscopy}

Spectroscopic follow-up was performed using grism \#13 on the ESO Faint Object Spectrograph and Camera 2 (EFOSC2) on the NTT over the nights of 2017 October 21-23 UT and 7-9 January 2018 (Program ID: 0100.A-0297, PI: Anguita) providing a resolution of $\sim$5.5\AA \ per pixel (with 2$\times$2 binning read-out). The Echellette Spectrograph and Imager \citep[ESI,][]{sheinis2002} on Keck 2 on the nights of 2017  November 18-19 UT was also used, in the default Echellette mode providing a resolution of $\sim$0.35 \AA \ per pixel at the central orders (Program ID:   2017B\_U110, PI: Fassnacht).

\begin{figure*}
	\includegraphics[width=2\columnwidth]{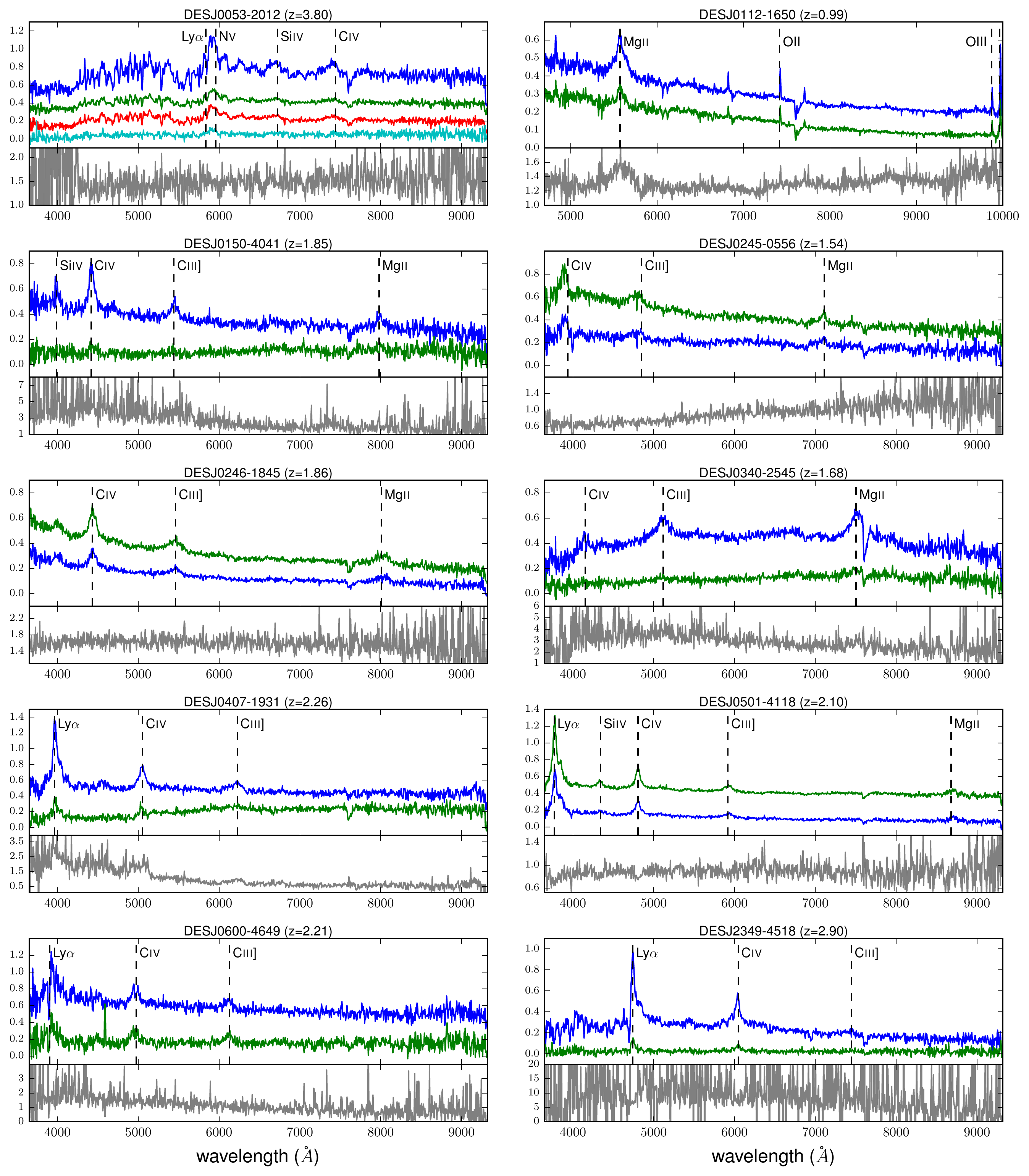}
    \caption{Spatially resolved spectra of quasar images of the confirmed lensed quasars. The flux is in arbitary units and in some cases the spectra have been offset to aid comparison of the spectra. Beneath each panel is a flux ratio against wavelength plot to aid comparison of spectra. For DESJ0053-2012 the flux ratio is between images A and B.}
    \label{fig:lens_spectra}
\end{figure*}

\begin{figure*}
	\includegraphics[width=2\columnwidth]{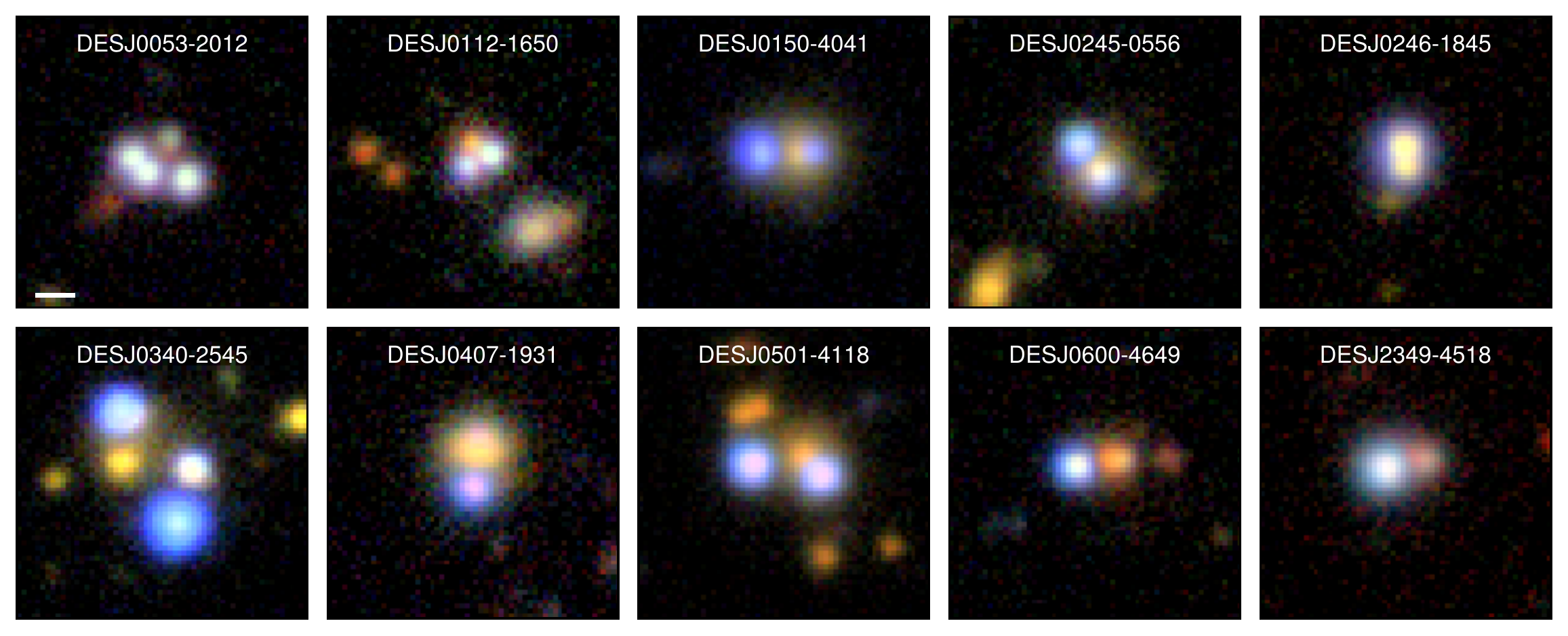}
    \caption{DES \textit{gri} colour images of the confirmed lensed quasars. Cutouts are 16.2 arcseconds on the side. The white scale bar in the first panel is 2 arcseconds. North is up, East is left.}
    \label{fig:lens_gris}
\end{figure*}

\begin{figure*}
	\includegraphics[width=2\columnwidth]{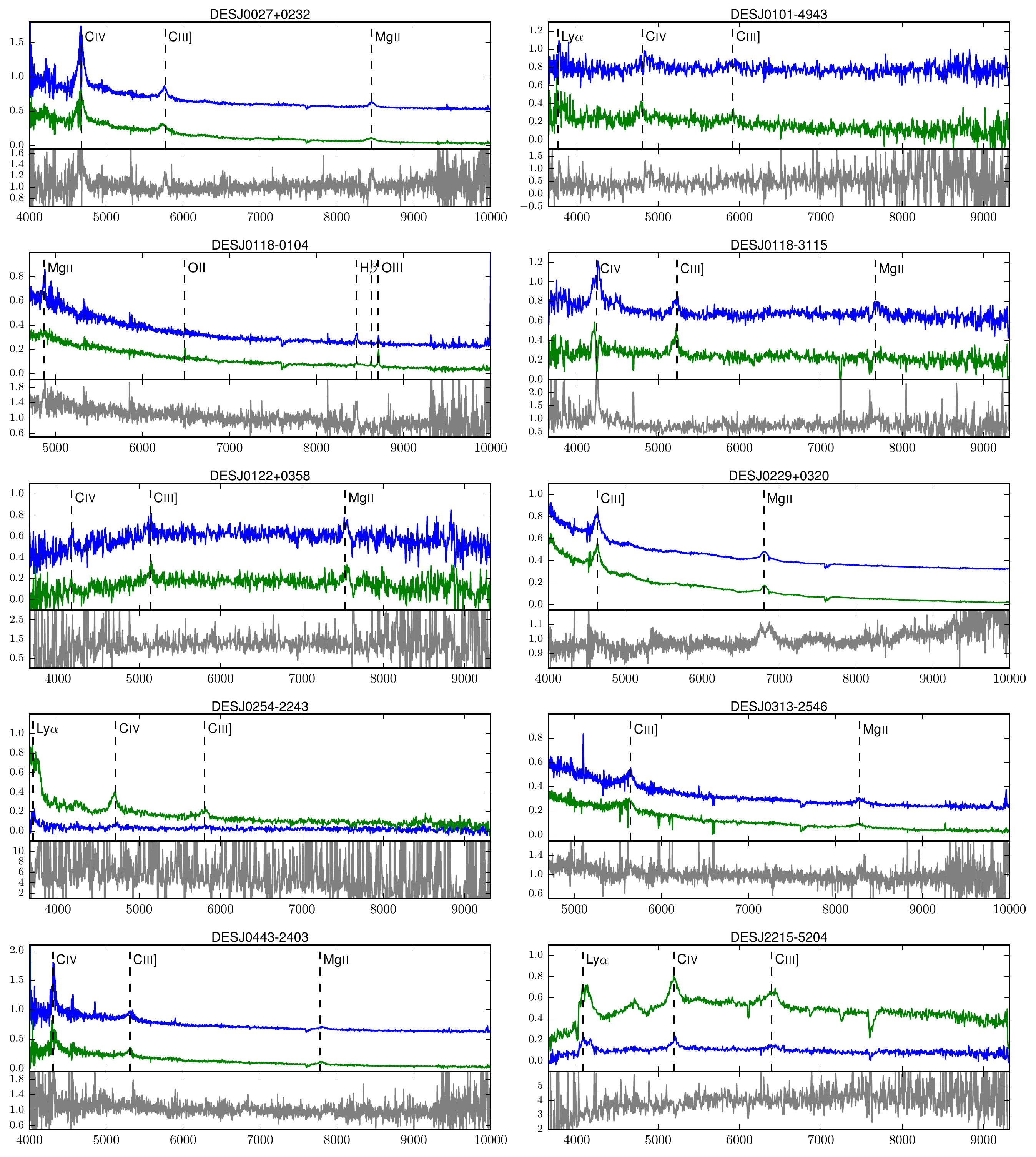}
    \caption{Spatially resolved spectra of pairs of quasars at similar (or possibly the same) redshifts. The flux is in arbitary units and in some cases the spectra have been offset to aid comparison of the spectra. Beneath each panel is a flux ratio against wavelength plot to aid comparison of spectra.}
    \label{fig:niqs_spectra}
\end{figure*}

All 2D background-subtracted spectra were visually inspected to confirm broad emission lines (or lack thereof) in the multiple spatially resolved components, and 1D spectra were extracted using Gaussian apertures of widths dependent on the seeing. Spectra of confirmed lensed quasars are shown in Figure \ref{fig:lens_spectra} (with their DES \textit{gri} colour cutouts shown in Figure \ref{fig:lens_gris}), and spectra of quasar pairs in Figure \ref{fig:niqs_spectra}.

\subsection{High-resolution imaging}
4 systems were observed using the Near InfraRed Camera 2 (NIRC2) on Keck-2 (PI: Treu, Program ID: U049), and 33 observed using the Southern Astrophysical Research Telescope (SOAR) Adaptive optics Module Imager (SAMI, PIs: Motta and Treu, programs: 1005 and 0138 respectively), as listed in Tables \ref{tab:lenses}, \ref{tab:qsopairs}, \ref{tab:inconc}, and \ref{tab:contaminants}. The data reduction and modelling for the two datasets are described as follows.

\begin{table*}
	\centering
	\caption{Confirmed lensed quasars. Selection: G1: \textit{Gaia} 1, G2: \textit{Gaia} 2, V: Variability, C: component fitting.}
	\label{tab:lenses}
	\begin{tabular}{lcccccl} 
		\hline
		Name & R.A. (J2000) & Dec. (J2000) & Selection & spectrum & imaging & outcome\\
		\hline
        DESJ0053-2012 & 13.4353	& -20.2092 & G1 & EFOSC2 & SOAR & \textbf{quad}, $z\approx$3.80\\
        DESJ0112-1650 & 18.1412 & -16.8410 & G1 & ESI & - & \textbf{double}, $z=$0.54 and $z=$0.99\\
        DESJ0150-4041 & 27.7369	& -40.6956 & G1, G2 & EFOSC2 & SOAR & \textbf{double}, $z=$1.85\\
        DESJ0245-0556 & 41.3565	& -5.9501 & G1, G2 & EFOSC & SOAR/NIRC2 & \textbf{double}, $z=$1.54 \\
        DESJ0246-1845 & 41.55083	& -18.7514 & C, G2 & EFOSC2 & SOAR/NIRC2 & \textbf{double}, $z=$1.86\\
        DESJ0340-2545 & 55.0351	& -25.7610 & G1 & EFOSC2 & SOAR/NIRC2 & \textbf{triple}, $z=$1.68\\
        DESJ0407-1931 & 61.9741 & -19.5225 & G1 & EFOSC2 & SOAR & \textbf{double}, $z=$0.288 and $z=$2.26\\
	    DESJ0501-4118 & 75.4413 & -41.3003 & G1 & EFOSC2 & SOAR & \textbf{double}, $z=$2.10\\
        DESJ0600-4649 & 90.1242 & -46.8168 & G1 & EFOSC2 & SOAR & \textbf{double}, $z=$2.21\\
        DESJ2349-4518 & 357.4924 & -45.3147 & G1 & EFOSC2 & - & \textbf{double}, $z=$2.90 \\
		\hline
	\end{tabular}
\end{table*}

\subsubsection{NIRC2}
\begin{figure*}
	\includegraphics[width=2\columnwidth]{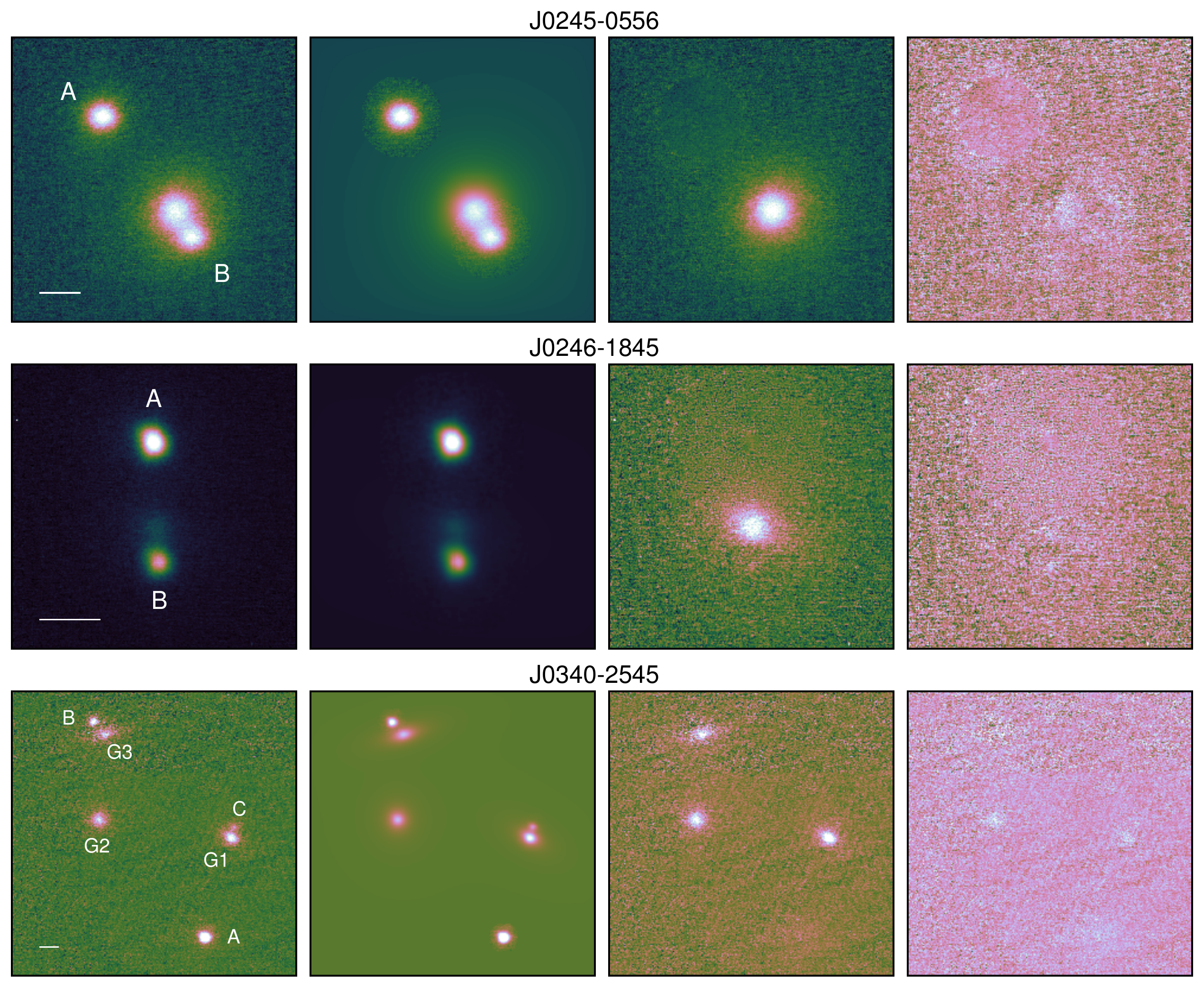}
    \caption{NIRC2 AO data for confirmed lensed quasars. From left to right: data, model, PSFs subtracted, and PSFs and galaxy subtracted. The white scale bar is 0.5 arcseconds, and labels refer to sources listed in Table \ref{tab:astrophotometry}. North is up, East is left. Flux is displayed using the cubehelix colour scheme \citep{cubehelix}.}
    \label{fig:NIRC2}
\end{figure*}

DESJ0245-0556, DESJ0246-1845, DESJ0340-2545, and DESJ0508-2748 were observed with the NIRC2 narrow camera, giving a 10$\times$10 arcsecond$^{2}$ field of view and 10 mas pixels. Observations were taken in the K band in order to maximise AO correction. Since there are no other stars to estimate the PSF in the field of the narrow camera, we reconstruct the PSF based on the data. While this can lead to fitting real structures such as host galaxy light---particularly degenerate for doubles---using analytical profiles often leaves significant residuals at the cores of PSFs where we would expect the lensed host galaxy to be brightest \citep{rusu2016, chen2016}. The PSF reconstruction is performed for each set of position and galaxy parameters, for a square PSF grid (with pixel sizes the same as the data) and linear interpolation, as described by \citet{ostrovski2018}. The reconstructed PSF from the best-fit model is used for convolution with the Sersic galaxy profiles. Since this PSF might not represent the true PSF, and due to atmospheric variations between frames, we include a positional uncertainty of 5 mas (half a pixel) in quadrature on our sampled statistical uncertainty. DESJ0508-2748 is fit to the noise with 2 PSFs and since it has a blended spectrum in both ESI and EFOSC2 data we label the system as inconclusive (Section \ref{J0508}). The other three candidates observed by NIRC2 are lensed quasars, with their data, PSF subtractions, and model subtractions shown in Figure \ref{fig:NIRC2}.

\subsubsection{SOAR}
33 candidates were observed with the SAMI instrument with its AO system SAM \citep{tokovinin2016}. Imaging was carried out in the \textit{z} band to maximise AO correction and optimise the contrast between quasar images and possible lensing galaxies. The pixel scale was 0.09 arcsec per pixel (2$\times$2 binning of 0.045 arcsec per pixel) and the typical exposure time was 3$\times$180 s.

\begin{figure}
	\includegraphics[width=\columnwidth]{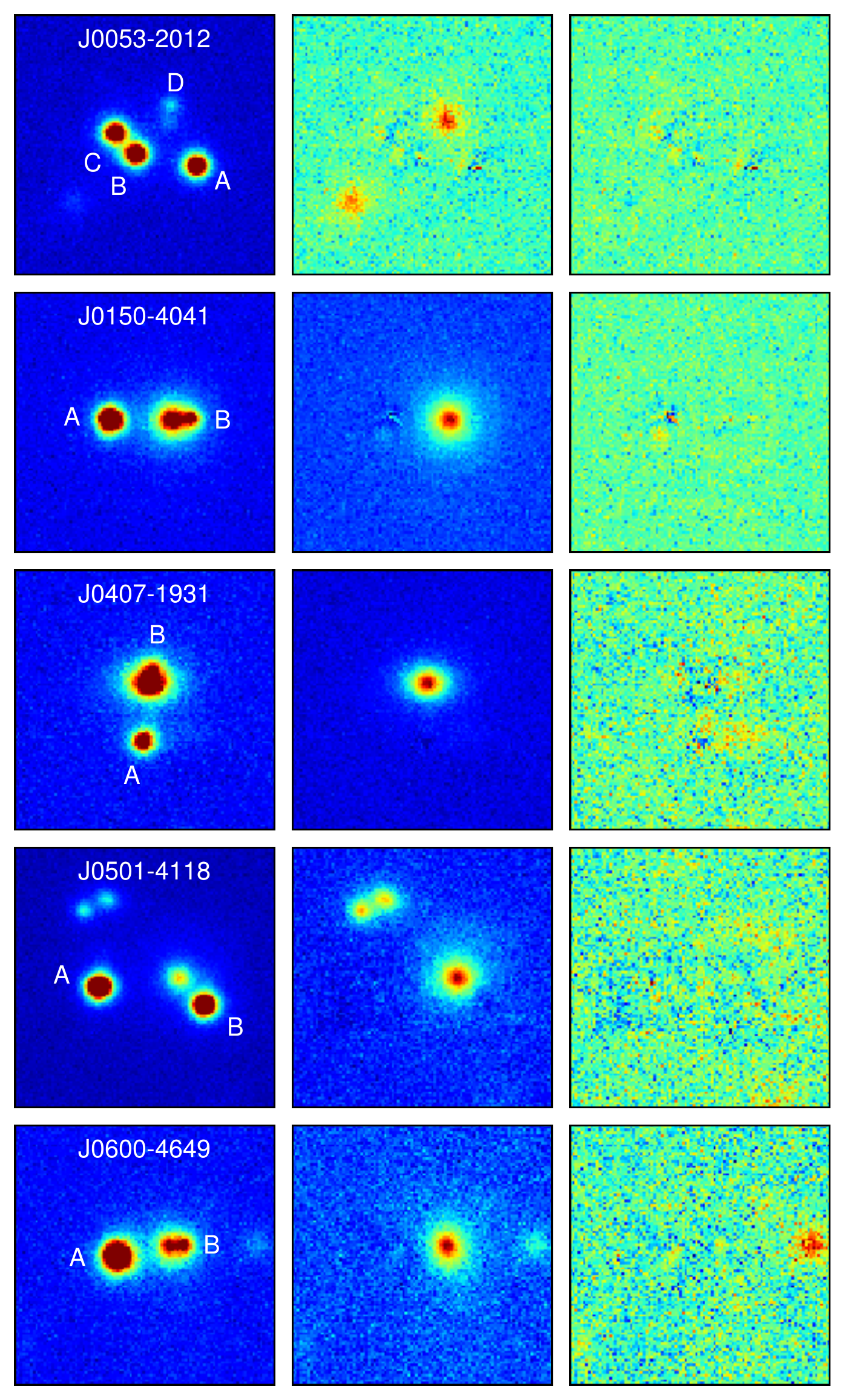}
    \caption{Left: SOAR \textit{z}-band cutouts for confirmed lensed quasars; middle: PSF subtracted images; right: PSF and galaxies subtracted. Cutouts are 9 arcseconds on the side. North is up, East is left.}
    \label{fig:SOAR}
\end{figure}

Given the large field of view (3$\times$3 arcminutes$^{2}$), nearby stars were used to fit Moffat profiles. When a good fit was achieved the Moffat parameters were used for the PSF model of the candidate system. If this PSF left visually obvious residuals, the Moffat parameters were included as part of the modelling of the system and simultaneously inferred with the galaxy and image parameters. If this was still left significant residuals, a nearby star was used, and pixel shifts were computed via a spline interpolation. All quasar pairs and inconclusive candidates, as listed in Tables \ref{tab:qsopairs} and \ref{tab:inconc} respectively, were consistent with 2 PSFs when modelling the SOAR data (or as a PSF and galaxy in the case of DESJ0402-4220). The data, PSF subtractions, and model subtractions for lenses without NIRC2 data are shown in Figure \ref{fig:SOAR}.

\subsection{Lens Modelling}
Given the measured image and galaxy positions, we are able to construct simple mass models for our confirmed lenses. NIRC2 and SOAR positions are used for all lenses except for DESJ0112-1650 and DESJ2349-4518, which lack high-resolution imaging. For these systems we use the DES data, which are modelled as in Section \ref{variability}. For the 8 doubly imaged systems, we use a singular isothermal ellipsoid \citep[SIE, ][]{kormann1994} to understand the basic properties of the system, such as Einstein radius and magnification. However, the inferred mass flattening is highly degenerate with the external shear, hence we set the latter to zero. Furthermore, we must use the flux ratio of the quasar images so we are not overfitting the data: 7 measurements (quasar image positions, galaxy position, and flux ratio) are used to fit 7 parameters (source position, galaxy position, Einstein radius, and mass axis ratio and position angle). Since these optical flux ratios may be strongly affected by microlensing or variability over the time-delay, we increase their uncertainties by 20 per cent in quadrature with the sampled statistical uncertainties. As there are zero degrees of freedom, we expect ${\chi}^{2}$ values of zero. Any discrepancy from this implies that the model is not a good description of the data. This might be caused by a complex mass distribution, or flux ratio anomalies due to substructure or microlensing, and so we report the ${\chi}^{2}$ contributions from galaxy position, image positions, and flux ratios separately for each lens system in Table \ref{tab:massmodels}, alongside the mass model parameters. Due to the simpliticity of a single SIE, our models can only hint at some true physical properties of the lens system, such as magnification, or whether a large shear is expected for the system due to a large mismatch between the modelled mass position angle and axis ratio and that of the light, which has been shown to be similar for many systems with shear properly accounted for \citep{shajib2019}. For the triply imaged system, DESJ0340-2545, and the quad, DESJ0053-2012, we explore more complex mass models as described in Section \ref{indivobjects}. 

\begin{table*}
	\centering
	\caption{Quasar Pairs. Selection: G1: \textit{Gaia} 1, G2: \textit{Gaia} 2, V: Variability, C: component fitting. NIQ stands for nearly identical quasar pair}
	\label{tab:qsopairs}
	\begin{tabular}{lcccccl} 
		\hline
		Name & R.A. (J2000) & Dec. (J2000) & Selection & spectrum & imaging & outcome\\
		\hline
        DESJ0027+0232 & 6.7619 & 2.5375 & G1 & ESI & - & $z=$2.02 NIQ \\
        DESJ0101-4943 & 15.3366 & -49.7234 & G1 & EFOSC2 & - & $z=$2.10 NIQ \\
        DESJ0118-0104 & 19.5501 & -1.0785 & G1 & ESI & - & $z=$0.74 NIQ \\
        DESJ0118-3115 & 19.6679 & -31.2619 & C &  EFOSC2 & SOAR & $z=$1.74 NIQ \\
        DESJ0122+0358 & 20.5990 & 3.9771 & G1 & EFOSC2 & SOAR & $z=$1.69 NIQ \\
        DESJ0152-4415 & 28.1183 & -44.2525 & C &  EFOSC2 & SOAR & projected QSOs, $z=$0.62 and $z=$2.17 \\
        DESJ0229+0320 & 37.4924 & 3.3419 & V & ESI & - & $z=$1.43 NIQ \\
        DESJ0254-2243 & 43.5720 & -22.7315 & G1 & EFOSC2 & SOAR & $z=$2.04 NIQ \\
        DESJ0313-2546 & 48.4088 & -25.7751 & G1 & ESI & SOAR & $z=$1.955 NIQ \\
        DESJ0330-4413 & 52.5070	& -44.2266 & G1 & EFOSC2 & SOAR & projected QSOs, $z=$0.52 and $z=$1.25\\
        DESJ0443-2403 & 70.9802 & -24.0572 & G1 & ESI & SOAR & $z=$1.78 NIQ \\
        DESJ2215-5204 & 333.9171 & -52.0679 & G1 & EFOSC2 & - & $z=$2.35 NIQ\\
		\hline
	\end{tabular}
\end{table*}

\begin{table*}
	\centering
	\caption{Median parameter values with 1$\sigma$ uncertainties for mass model and galaxy light profiles of confirmed lensed quasars. \textit{b}=Einstein radius, \textit{PA}= position angle (East of North), \textit{q} = axis ratio, and $\mu$ = magnification. Given the group lensing DESJ0340-2545, more detailed mass models are described in the text and Table \ref{tab:J0340_massmodels} for this system. The model we use for DESJ0053-2012 is a 2 SIE model, as described in the text. For this model,  we expect a good fit to have ${{\chi}^2}\approx$ 3.}
	\label{tab:massmodels}
	\begin{tabular}{cccccccc}
		\hline
		name & $b$ ($\arcsec$) & $PA_\textrm{SIE}$ &$q_\textrm{SIE}$ & $PA_\textrm{phot}$ & $q_\textrm{phot}$& ${{\chi}^2}_{\textrm{gal., images, flux}}$ & ${\mu}_{\textrm{total}}$ (${\mu}_{\textrm{individual}}$)  \\
		\hline
        \noalign{\vskip 0.7mm}
        DESJ0053-2012 & $1.17_{0.01}^{0.01}$ & $168_{3}^{2}$ & $0.56_{0.03}^{0.04}$ & $21_{9}^{10}$ & $0.67_{0.09}^{0.11}$ & 0.19, 0.03, 2.21 & $17.1_{1.4}^{0.9}$ ($6.4, 5.8, 4.3, 0.5$)\\
        \noalign{\vskip 0.7mm}
        DESJ0112-1650 & $0.77_{0.01}^{0.01}$ & $165_{5}^{6}$ & $0.69_{0.02}^{0.01}$ & $174_{2}^{1}$ & $0.77_{0.02}^{0.01}$ & 0.00, 0.00, 0.00 & $5.64_{0.31}^{0.31}$ ($3.65, 2.01$)\\
        \noalign{\vskip 0.7mm}
        DESJ0150-4041 & $1.44_{0.02}^{0.02}$ & $104_{4}^{6}$ & $0.86_{0.05}^{0.04}$ & $90_{14}^{13}$ & $0.96_{0.02}^{0.02}$ & 0.00, 0.00, 0.00 & $4.29_{0.23}^{0.66}$ ($3.47, 0.82$)\\
        \noalign{\vskip 0.7mm}
        DESJ0245-0556 & $0.90_{0.01}^{0.02}$ & $131_{2}^{3}$ & $0.67_{0.10}^{0.11}$ & $157_{3}^{3}$ & $0.947_{0.005}^{0.005}$ & 0.00, 0.00, 0.00 & $3.15_{0.06}^{0.09}$ ($1.96, 1.16$)\\
        \noalign{\vskip 0.7mm}
        DESJ0246-1845 & $0.49_{0.01}^{0.01}$ & $69_{28}^{10}$ & $0.92_{0.05}^{0.03}$ & $78_{1}^{1}$ & $0.44_{0.03}^{0.02}$ & 0.00, 0.00, 0.00 & $4.61_{0.11}^{0.17}$ ($3.11, 1.50$)\\
        \noalign{\vskip 0.7mm}
        DESJ0407-1931 & $1.30_{0.01}^{0.02}$ & $134_{20}^{20}$ & $0.92_{0.03}^{0.02}$ & $87_{1}^{1}$ & $0.70_{0.02}^{0.02}$ & 0.00, 0.00, 0.00 & $3.55_{0.11}^{0.22}$ ($2.78, 0.76$)\\
        \noalign{\vskip 0.7mm}
        DESJ0501-4118 & $1.97_{0.03}^{0.03}$ & $14_{3}^{3}$ & $0.43_{0.02}^{0.02}$ & $41_{7}^{7}$ & $0.90_{0.02}^{0.03}$ & 0.00, 0.00, 0.02 & $3.36_{0.19}^{0.22}$ ($1.48, 1.88$)\\
        \noalign{\vskip 0.7mm}
        DESJ0600-4649 & $1.22_{0.02}^{0.02}$ & $87_{10}^{6}$ & $0.86_{0.07}^{0.05}$ & $12_{6}^{6}$ & $0.85_{0.04}^{0.04}$ & 0.00, 0.00, 0.00 & $3.90_{0.49}^{0.25}$ ($3.22, 0.67$)\\
        \noalign{\vskip 0.7mm}
        DESJ2349-4518 & $1.24_{0.02}^{0.02}$ & $69_{3}^{2}$ & $0.59_{0.03}^{0.03}$ & $57_{11}^{9}$ & $0.73_{0.06}^{0.06}$ & 0.26, 0.01, 0.98 & $4.84_{0.50}^{0.59}$ ($4.10, 0.83$)\\
        \noalign{\vskip 0.7mm}
        \hline
	\end{tabular}
\end{table*}

\subsection{Notes on Individual Objects} \label{indivobjects}
\subsubsection{DESJ0053-2012}
This system is the only confirmed quad from the 2017/2018 follow-up campaign. It was selected by the \textit{Gaia}-\textit{WISE} selection of Section \ref{gaia1} as a \textit{Gaia} DR1 double (G=19.21, 19.43) corresponding to a red \textit{WISE} detection (\textit{W1}$-$\textit{W2}=0.55). The redshift of the system is $z\approx$3.8. It is not more precisely determined due to the strong absorption of Lyman alpha.

\begin{figure*}
	\includegraphics[width=2\columnwidth]{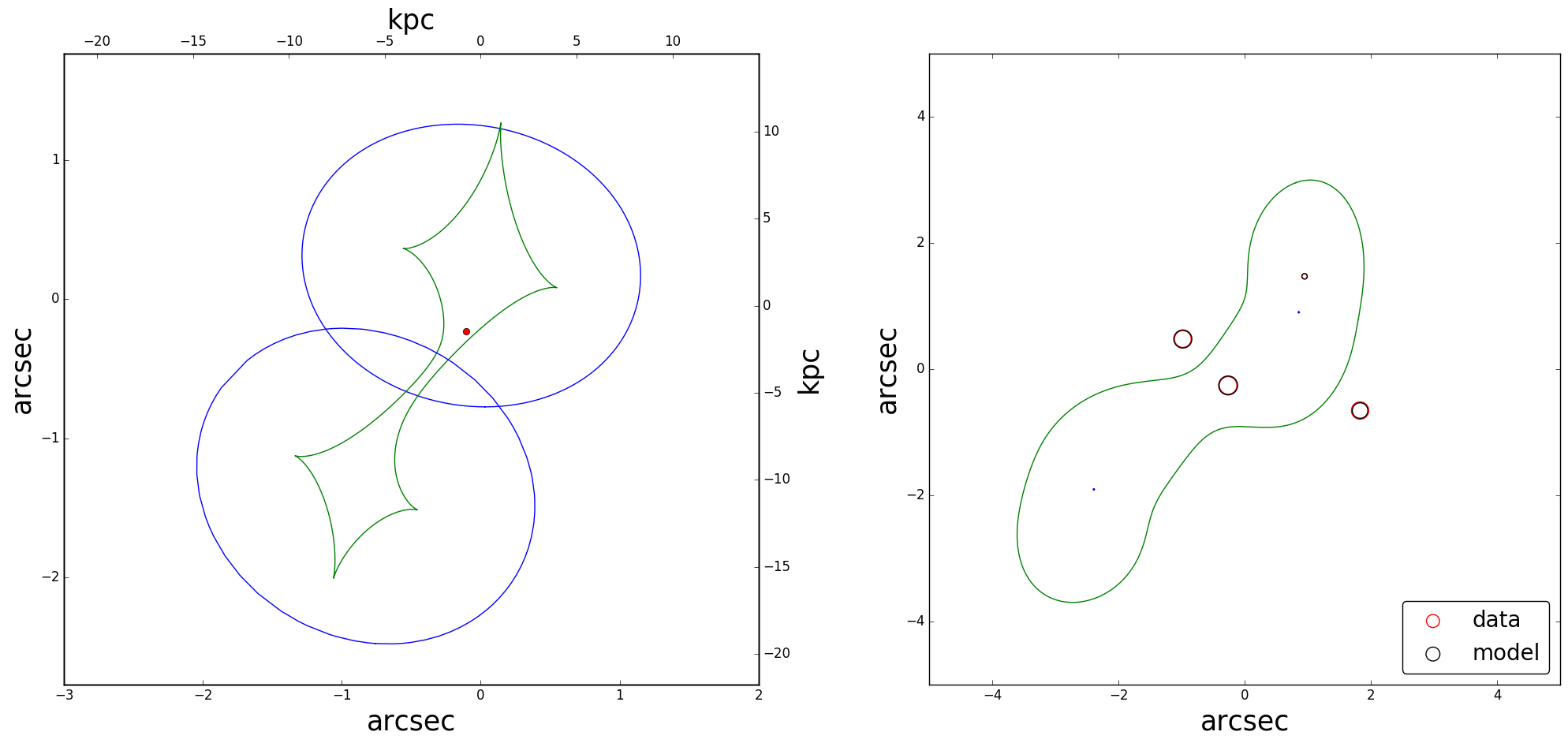}
    \caption{DESJ0053-2012 best-fit model. \textit{Left}: source plane caustics with source position overlaid in red. \textit{Right}: image plane critical curves with measured image positions, and best-fit model image positions overlaid, with area representing flux. North is up, East is left.}
    \label{fig:J0053_curves}
\end{figure*}

A SIE+shear model is insufficient to reproduce the positions and flux ratios, providing a best-fit ${{\chi}^2}\approx30$ for 4 degrees of freedom. The model requires a strong shear of 0.22, 159 degrees East of North, while the companion galaxy G2 is 4.3 arcseconds 131 degrees East of North. We choose to explicitly model G2 as an SIE with shape fixed to that of the light (as measured in the $z$-band SOAR data), and leave the Einstein radius as a free parameter. This new model provides a best-fit ${{\chi}^2}=$ 2.43 for 3 degrees of freedom, now requiring a shear of 0.13, 20 degrees East of North. The caustics and critical curves of this model are shown in Figure \ref{fig:J0053_curves}. The main contribution to the ${{\chi}^2}$ is from the flux ratios, with B being 25 per cent fainter than predicted. This is consistent with the expectations of microlensing suppression since this image is a saddle point \citep{schechter2002}. Finally, we note that, assuming a lens redshift of 0.7, expected time delays are 22, 26, and 137 days, in the expected image ordering ACBD.

\subsubsection{DESJ0112-1650}
\begin{figure*}
	\includegraphics[width=2\columnwidth]{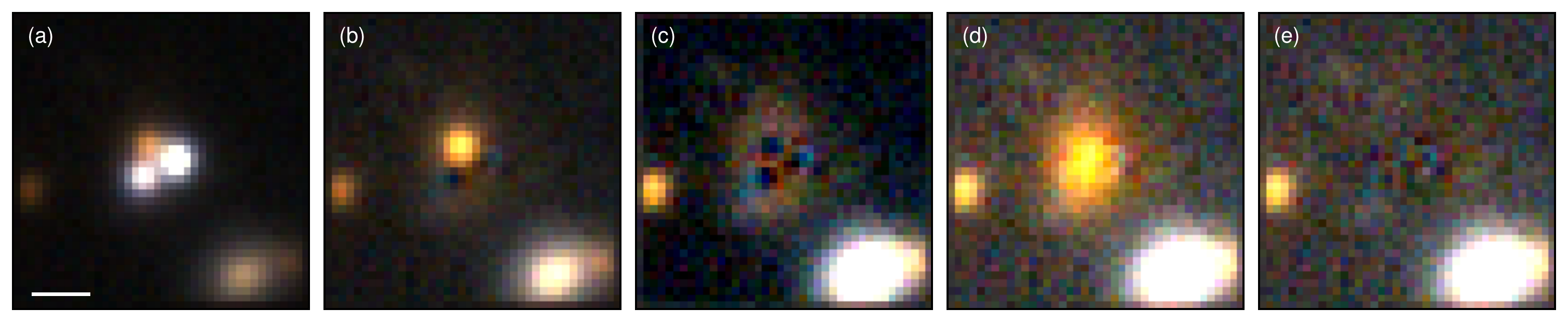}
    \caption{J0112-1650 \textit{gri} stacked models from single-epoch images. (a) data, (b) three-PSF model subtracting the two blue PSFs, (c) three-PSF model subtracting all PSFs, which clearly shows an extended component in the residuals, (d) three-PSF and galaxy model subtracting the three PSFs, and (e) three-PSF and galaxy model subtracting all components. The white scale bar is 2 arcseconds.}
    \label{fig:J0112}
\end{figure*}

Resolved ESI spectra show two quasars with very similar spectra at $z=$0.99. There is absorption due to a massive galaxy seen in both quasar images at $z=$0.52 (from Ca H, K, G, and Na at 6063, 6116, 6635, and 9084 \AA \ respectively), which is a promising sign of a lensing galaxy. While we lack high-resolution imaging of this system, the DES data clearly shows a redder object offset from the line joining the two blue point sources. This is often seen with fold quad configurations, where the faint counterimage is either highly reddened or the data are not deep enough, while one of the observed PSFs is a pair of close images, as was observed in the cases of PSJ0030-1525 \citep{lemon2018} or SDSSJ1330+1810 \citep{oguri1330}. We model the system as two point sources, and subsequently as three point sources, which both show an extended galaxy in the residuals. Therefore we adopt three point sources and a galaxy (Sersic profile) as the fiducial pixel model for this system, as this fits the data without any obvious residual structures (Figure \ref{fig:J0112}). We currently cannot determine whether the third point source, C, is a highly reddened quasar image, a foreground star, or structure due to the lensing galaxy. Our mass model given in Table \ref{tab:massmodels} assumes only A and B are quasar images, with G being the only lensing galaxy. The mean reduced ${\chi}^{2}$ values for A, B, and C from multi-epoch modelling (as described in Section \ref{variability}) are 6.72, 5.42, and 2.74, respectively. However, this does not imply C is necessarily less variable, since its photometric uncertainties are larger relative to the brighter components, A and B. The value is consistent with those from stars in quasar+star systems and should not be taken as a detection of variability. High-resolution imaging is required to clarify the nature of this system.

\subsubsection{DESJ0229+0320}
This system consists of two quasars at a redshift of 1.43, separated by 2.13 arcseconds. The spectra are similar, and the multi-band multi-epoch modelling (Section \ref{variability}) shows similar long-term variability. Furthermore, the apparent brightnesses of these quasars (\textit{Gaia} magnitudes of 18.15 and 18.79) place them at the bright end of the luminosity function for quasars. These characteristics are all in accordance with strong gravitational lensing --- similar variations, similar spectra, and large apparent brightness due to magnification --- however no lensing galaxy is seen in the deep DES images. To investigate how faint a lens galaxy can be for such a system, we find the $i$-band magnitudes of a subset of the \citet{om10} mocks, namely those with similar redshift sources ($z=$1.43$\pm$0.2) and  similar image separations of 2.13$\pm$0.2 arcseconds. Of the 192 systems satisfying theses criteria, the faintest lensing-galaxy magnitude is 20.8 ($z\sim$ 0.9; expected once in 100,000 square degrees of sky). This magnitude is reached in a single-epoch image \citep[for a S/N ratio of 10 for an isolated point source,][]{desdr1}; however, if it lies close to one of the quasar images, or is particularly extended, then it might not be apparent after PSF subtraction. However, modelling 0.5$^{\prime \prime}$-seeing archival Hyper Suprime Cam $z$-band data of the system also reveals no evidence of a lensing galaxy. We believe this is a binary quasar, but high-resolution imaging is required to fully rule out the lensing hypothesis.

\subsubsection{DESJ0340-2545}
This system was found through a \textit{Gaia} and \textit{WISE} selection following Section \ref{gaia1}. A slit positioned at 27 degrees East of North confirms two quasar images separated by 6.8 arcseconds, each at $z=$1.68 (see Figure \ref{fig:lens_gris}). NIRC2 imaging of the system clearly shows three lensing galaxies, two quasar images, and one further object, north of G1 (as labelled in Figure \ref{fig:NIRC2}). Overlaying the \textit{Gaia} detections for the system on the NIRC2 data reveals that this further object is exactly centred on a \textit{Gaia} optical detection, and the DES colour image reveals a blue object blended with G1. We investigate whether this object could be the third image of the system, as these faint central images are often observed in lenses with multiple galaxies, resulting in an image slightly offset from one of the lensing-galaxy centres, for example, PSJ0630-1201 \citep{ostrovski2018}. To investigate whether C is another quasar image, we create lens models based only on the confirmed quasar images, and see whether the best-fitting lens models naturally predict another image near C. However, we are severely underconstrained given the complex mass distribution of the galaxy group and only two quasar positions and their, often untrustworthy, flux ratio. Given that our model will require a source position, we are left with only three degrees of freedom for our lens model. We choose to model the mass contributions of the three galaxies, G1, G2, and G3, set as singular isothermal ellipsoids, with their flattening parameters set to those of the light (Table \ref{tab:J0340_parameters}), and pinned to the measured light positions. Their Einstein radii are all modelled by 1 parameter, $b$, assuming a constant mass-to-light ratio amongst the galaxies, i.e. using the galaxy flux ratios to set the Einstein radii ratios ($L \propto M \propto b^{0.5}$). We also choose to include an external shear so we have no degrees of freedom remaining, and expect a good fit to have ${\chi}^{2}\approx$ 0. This model reproduces the two bright confirmed quasar images and their flux ratio well. It also predicts a third image 0.22 arcseconds away from the postulated image C, and with a flux 15 per cent that of A, while C has a measured flux of 10 per cent of A. This prediction is enough for us to consider C as a third quasar image without its spectroscopic confirmation. Further evidence for C being a quasar image comes from the remarkably similar flux ratios of the point sources in \textit{Gaia} and in the NIRC2 K-band data (as given in Table \ref{tab:J0340_parameters}), suggesting they have similar spectral energy distributions (SEDs). We can use the three extra constraints this image provides to consider more complex lens models. Considering the same model as before, but now including C as a required quasar image to be reproduced, produces a poor best-fit with $\chi^2$ = 676, given 3 degrees of freedom. We consider two further lens models. Firstly, we allow the mass-to-light ratio to vary between the lensing galaxies, i.e. fit for the Einstein radius of each galaxy. This results in a good fit to the astrometry, but image A is predicted to be about as bright as B, while it is in fact measured to be 2.7 times brighter. This model might describe the system well, considering there could be large variability over the time delay, so later DES and LSST observations will help exclude this possible model. Our final model fixes the Einstein radii as in our initial models, but now includes an SIS, representing a dark matter halo shared by the lensing galaxies. Its best-fit position is $\Delta x$, $\Delta y$ = 0.75, 0.60 arcseconds from A (see Table \ref{tab:astrophotometry}), and the $\chi^2$ for this model is 0.43, given 0 degrees of freedom. Since the true mass distribution within this group is likely very complicated, these models should only serve as a discussion of the possible probes this system could offer, in particular if deeper high-resolution imaging is pursued to reveal the multiply imaged host-galaxy arcs. 

\begin{figure}
	\includegraphics[width=\columnwidth]{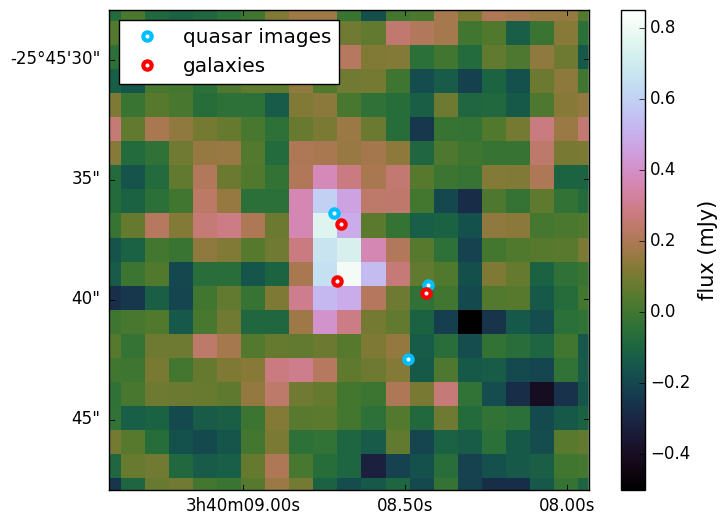}
    \caption{VLASS 3GHz image of J0340-2545 with quasar and galaxy positions overlaid, as measured from the NIRC2 data and shifted to the \textit{Gaia} DR2 frame. North is up, East is left.}
    \label{fig:J0340_VLASS}
\end{figure}

DESJ0340-2545 has a 15 mJy detection by the VLA Sky Survey (VLASS) at 3 GHz \citep{VLASS} and a 2.6 mJy detection in NVSS at 1.4GHz \citep{nvss1998}. Given the 2.5 arcsecond resolution of VLASS, and the large separation of this system, we are able to determine whether the radio emission is due to the quasar images. Figure \ref{fig:J0340_VLASS} shows the VLASS 3GHz cutout with \textit{Gaia} detections overlaid. The emission is consistent with coming from one or two of the lensing galaxies, G2 and G3 as labelled in Figure \ref{fig:J0340_VLASS}.

\begin{table}
	\centering
	\caption{DESJ0340-2545 parameters based on NIRC2 data.}
	\label{tab:J0340_parameters}
	\begin{tabular}{cccccl} 
		\hline
		Component & q & PA & K'(\textit{Gaia} G) flux ratio\\
		\hline
        A & --- & --- & 9.59 (9.55) \\
        B & --- & --- & 3.49 (3.50) \\
        C & --- & --- & 1.00 (1.00) \\
		\hline
        G1 & 0.54 $\pm$ 0.01 & 150 $\pm$ 1 & 1.00 (---) \\
        G2 & 0.92 $\pm$ 0.02 & 96 $\pm$ 7 & 0.95 (---) \\
        G3 & 0.344 $\pm$ 0.008 & 15.5 $\pm$ 0.5 & 0.97 (---) \\
		\hline
	\end{tabular}
\end{table}

\begin{table*}
	\centering
	\caption{DESJ0340-2545 mass models}
	\label{tab:J0340_massmodels}
	\begin{tabular}{ccccccc} 
		\hline
		Model & Fit C as an image? & Constraints & Model Parameters & $\chi^2$ astrometry/flux & total $\chi^2$ & $\mu$\\
		\hline
		1 & No & 5: $XY_{AB}$, $f_{AB}$ & 5: $XY_{S}$, $b$, $\gamma$, $\theta_\gamma$ & 0.00, 0.00 & 0.00 & 4.40 \\
		2 & Yes & 8: $XY_{ABC}$, $f_{ABC}$ & 5: $XY_{S}$, $b$, $\gamma$, $\theta_\gamma$ & 670.6, 5.3 & 675.9 & 4.53 \\
		3 & Yes & 8: $XY_{ABC}$, $f_{ABC}$ & 7: $XY_{S}$, $b_{1}$, $b_{2}$, $b_{3}$, $\gamma$, $\theta_\gamma$ & 0.02, 9.51 & 9.53 & 5.25 \\
		4 & Yes & 8: $XY_{ABC}$, $f_{ABC}$ & 8: $XY_{S}$, $b$, $\gamma$, $\theta_\gamma$, $XY_{HALO}$, $b_{HALO}$ & 0.01, 0.42 & 0.43 & 11.6 \\
		\hline
	\end{tabular}
\end{table*}

\subsubsection{DESJ0407-1931}
\begin{figure*}
	\includegraphics[width=2\columnwidth]{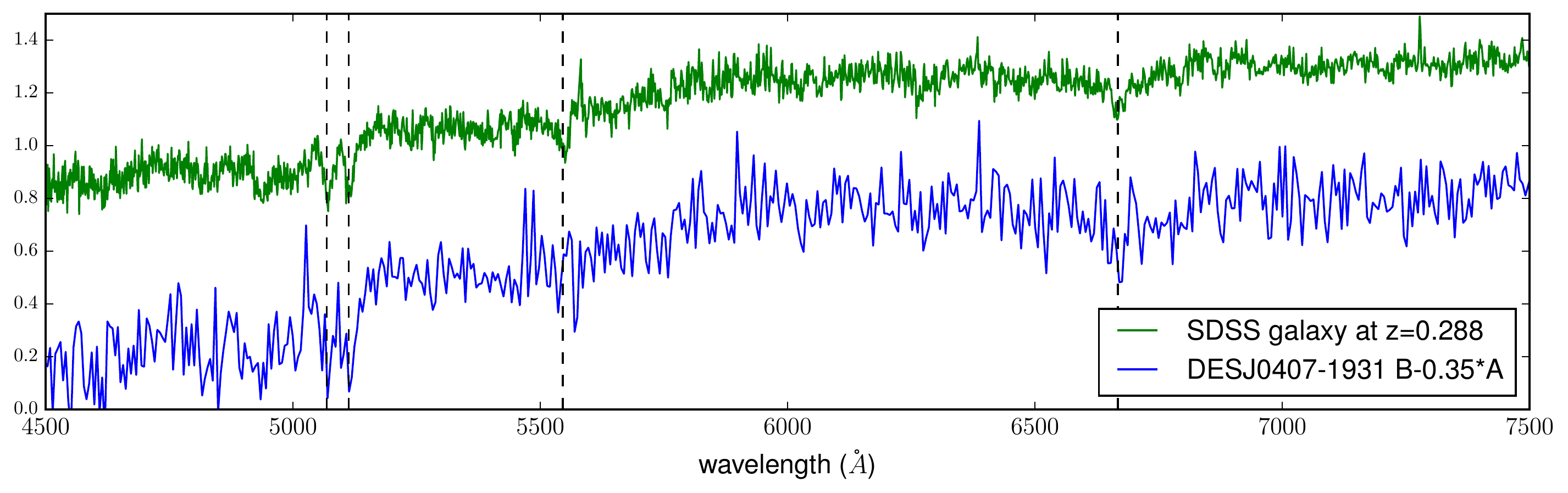}
    \caption{NTT spectrum of the lensing galaxy of DESJ0407-1931, created by subtracting a scaled spectrum of image A from the B+G blended spectrum. A high signal-to-noise spectrum from SDSS of a galaxy at $z=$0.288 is plotted for comparison, highlighting the Ca H, K, and Mg lines at 5070, 5115, and 6670\AA \, respectively.}
    \label{fig:J0407_spec}
\end{figure*}

The fainter image of this double is significantly blended with the lensing galaxy, but it is clearly detected in the DES stacked data, and the SOAR imaging. Resolved spectra also show identical emission lines of a quasar at $z=2.26$. By subtracting off the scaled NTT spectrum of A from B, we are able to see clear signs of the lensing galaxy absorption lines at $z=$0.288 as shown in Figure \ref{fig:J0407_spec}. The stacked modelled residuals show an excess to the East and West of the brighter image, potentially caused by the lensed host galaxy.

\subsubsection{DESJ0508-2748} \label{J0508}
The NTT and ESI spectra for this object both show a blended $z=$1.14 quasar without any other obvious features. NIRC2 data reveals two point sources, separated by 0.65 arcseconds, with no obvious lensing galaxy after PSF subtraction. We leave this object as inconclusive since the spectra are not resolved and there are no other indicators as to its nature (e.g., high proper motion \textit{Gaia} detection, additional spectral features, or detection of a single quasar host galaxy).

\subsubsection{DESJ2349-4518}
\begin{figure}
	\includegraphics[width=\columnwidth]{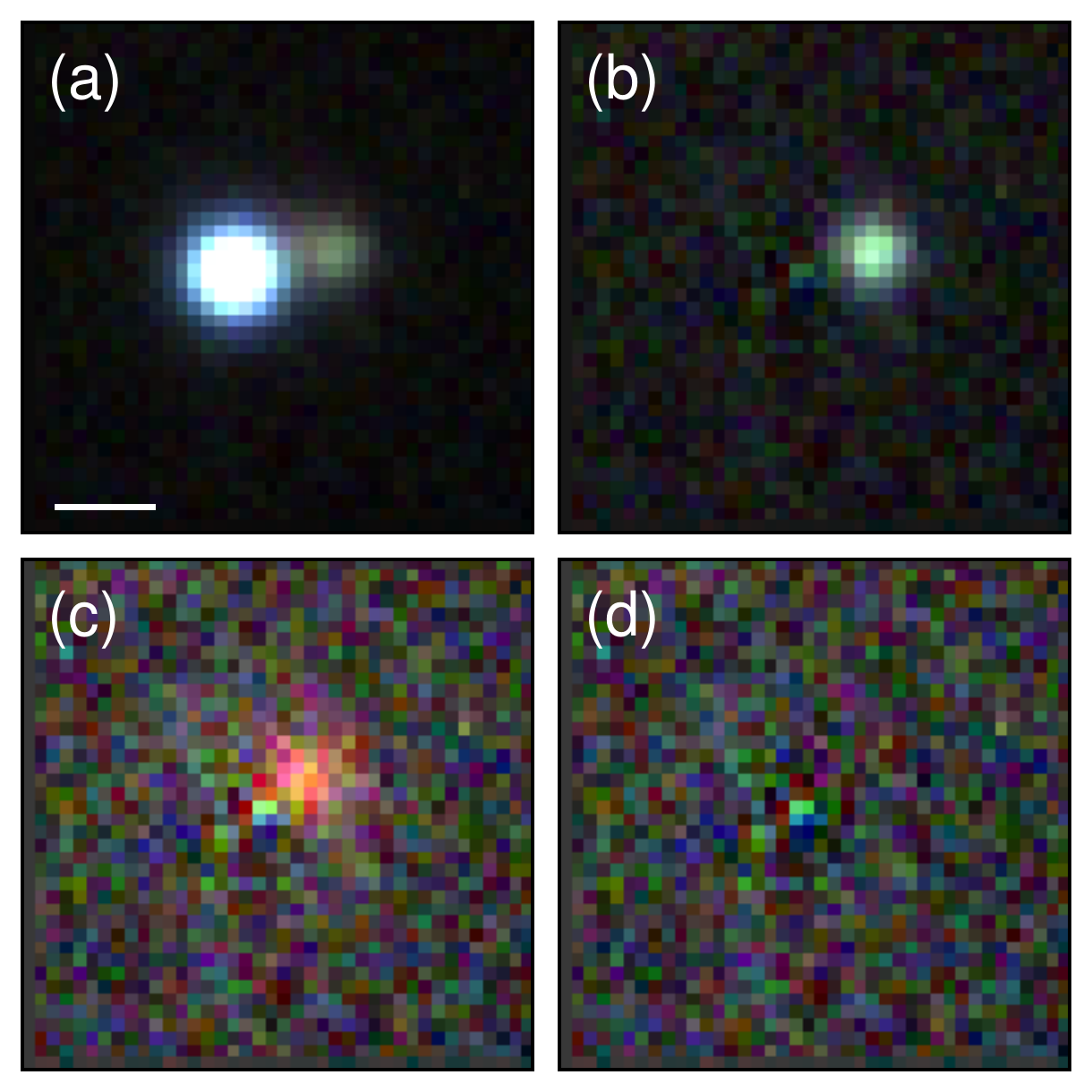}
    \caption{DES \textit{gri} stacked model for DESJ2349-4518; (a) data, (b) subtraction of image A and lensing galaxy, (c) subtraction of both quasar images, and (d) subtraction of all components. The white scale bar is 2 arcseconds.}
    \label{fig:J2349}
\end{figure}

This system is a high-flux-ratio double (7.5 to 1), with a SUMSS (Sydney University Molonglo Sky Survey) detection of 13.6 mJy at 843 MHz \citep{sumss}, with a source redshift of $z=2.90$. This lens system lacks SOAR or NIRC2 data, so we show DES \textit{gri} colour images with the brighter PSF and both PSFs subtracted in Figure \ref{fig:J2349}. This system is the only double with a non-zero ${\chi}^{2}$ for the single SIE fit. The largest contribution to the residuals is from the observed flux ratio being higher than the model predicts. This is most likely explained by microlensing, variability, or attenuation of the fainter image due to the lensing galaxy.

\subsection{Nearly Identical Quasar Pairs}
In Table \ref{tab:niqlens}, we present crude estimates for how faint a possible lensing galaxy can be for each of the 10 nearly identical quasar pairs (NIQs). The method is based only on separation and source redshift, following the prescription explained in \citet{strides1}, using the LensPop package \citep{collett2015}. We note that the DES median coadded catalog depths for a 1.95" diameter aperture at S/N = 10 in the \textit{griz} bands are 24.3, 24.1, 23.4, and 22.7mag respectively \citep{desdr1}. However, we cannot conclusively rule out the existence of lensing galaxies since they are extended (and hence are more difficult to detect) and our PSF subtraction may fit part of this lensing galaxy flux, especially if one image lies close to the lens and/or the systems have small angular separations.

\begin{table}
	\centering
	\caption{Estimates of the faintest magnitudes of possible lenses in \textit{griz} for the NIQs, based on image separation and source redshift.}
	\label{tab:niqlens}
	\begin{tabular}{lcccccc} 
		\hline
		Name & ${\Delta}{\theta}$ ($^{\prime \prime}$) & $z$ & \textit{g} & \textit{r} & \textit{i} & \textit{z} \\
		\hline
        DESJ0027+0232 & 2.72 & 2.02 & 26.3 & 24.0 & 22.2 & 20.9 \\
        DESJ0101-4943 & 4.02 & 2.10 & 25.5 & 22.7 & 21.4 & 20.1 \\
        DESJ0118-0104 & 1.74 & 0.74 & 22.9 & 21.0 & 19.9 & 19.4 \\
        DESJ0118-3115 & 1.00 & 1.74 & 28.1 & 25.3 & 24.0 & 22.7 \\
        DESJ0122+0358 & 1.71 & 1.69 & 26.9 & 24.0 & 22.8 & 21.5 \\
        DESJ0229+0320 & 2.08 & 1.43 & 25.8 & 23.0 & 21.7 & 20.6 \\
        DESJ0254-2243 & 2.32 & 2.04 & 26.7 & 24.4 & 22.6 & 21.3 \\
        DESJ0313-2546 & 2.20 & 1.96 & 26.7 & 24.3 & 22.6 & 21.3 \\
        DESJ0443-2403 & 1.85 & 1.78 & 26.9 & 24.1 & 22.8 & 21.4 \\
        DESJ2215-5204 & 2.74 & 2.35 & 26.5 & 25.0 & 22.8 & 21.3 \\
		\hline
	\end{tabular}
\end{table}

\section{Variability Modelling} \label{variability}
A central aspect of the Dark Energy Survey is its repeated observations of a large area of sky over 5 years. This allows us to extract variability information in the \textit{grizY} bandpasses (with central wavelengths 4827, 6432, 7827, 9179, 9900\AA \, respectively), as explored for efficient quasar detection in Section \ref{variabilityselection} \citep{flaugher2015}. Extracting variability information for the multiple components of nearby blends of objects offers a promising way to (i) remove quasar+star systems from future follow-up of potential doubly imaged quasars \citep{kochanek2006var}, and (ii) to prioritise possible lenses amongst nearly identical quasar pairs by looking for similar variability, as has been done with targeted repeat observations of SDSS candidates \citep{sergeyev2016, shalyapin2018}. For lensed quasars in high-cadence fields, constraints can be placed on the time delays \citep{zuzanna2018}. However, applying variability analysis on pipeline magnitudes of close-separation pairs will lead to spurious results due to flux-sharing \citep{tewes2013} and variable seeing, and excludes applicability to all cases of blended sources that are not segmented by the source extraction software.

This problem has been well-explored by teams dedicated to lensed quasar monitoring for extraction of lightcurves and hence time delays. In particular, photometry has been extracted through fitting multiple PSFs \citep[e.g.,][]{goicoechea2010, koptelova2010}, fitting multiple PSFs and parametric galaxy components \citep[e.g.,][]{kochanek2006, shalyapin2017, hainline2013}, difference imaging \citep[e.g.,][]{fohlmeister2013, giannini2017}, aperture photometry for wide-separation systems \citep[e.g.,][]{ovaldsen2003, dahle2015}, and non-parametric deconvolutional techniques \citep[e.g.,][]{burud2002, vuissoz2007, bonvin2018}.

\subsection{A photometry pipeline for lensed quasars} \label{pipeline}
We note that understanding the PSF is vital for deriving reliable photometry from each epoch. Tests using PSFEx \citep{bertin2011} reconstructions do not fit the PSFs of our systems precisely enough, and there is no knowledge of the uncertainty on the PSF model over which to marginalise. Using nearby stars or stacks of nearby stars as the PSF is also often inconsistent with the PSF of the system, perhaps due to spatial variations, colour mismatches, or brightness mismatches between the system PSF and the star PSFs, which can exhibit different shapes due to flux-dependent charge interactions \citep[i.e., the brighter-fatter effect,][]{gruen2015, walter2015}. We therefore opt for an initial PSF fit by a Moffat profile to a nearby star, but we later allow the data from each epoch to fit this profile, and marginalise over the Moffat parameters. For systems with bright point sources (\textit{Gaia} G$<$18), this profile often fails to describe the PSF to the noise as expected, and we remove the relevant frames from our analysis. For the remaining frames of these bright systems, the photometric uncertainty is often $<$0.01 mag, and so only a few frames are needed to statistically verify variability. Poor Moffat fits are uncommon for the fainter systems, for which more frames are required for a robust detection of variability given the increased photometric uncertainties on fainter point sources.

For the following, we use all single-epoch images with DES ``Final Cut" processing from the first four years of DES \citep{morganson2018}. The steps of the modelling pipeline are as follows:

\begin{enumerate}
\item 40$\times$40 pixel cutouts of all single-epoch images centred on the relevant system are inspected, and any showing artefacts/cosmic rays or significantly poor seeing (over 2 arcseconds FWHM based on the FITS header information), are excluded from the modelling.
\item To determine the zeropoint in each band, stellar objects on the same CCD are found by plotting $MAG\_AUTO-MAG\_PSF$ vs $MAG\_AUTO$ \citep[as derived from SExtractor,][]{bertin1996}, fitting a line to the stellar locus, and selecting all objects within 0.05 mag of this line and with catalogued magnitudes between 15.5 and 19. The sky background is estimated from a fit to the histogram of pixels between 5 and 25 arcseconds around each star, after a 5-$\sigma$ clip. The background level is determined through the Bayesian model of \citet{bijaoui1980}, however we also include some fraction of empty sky pixels in this model. Following the notation of \citet{bijaoui1980}, the (unnormalised) probability distribution of true sky flux ($i$) values is:
\begin{equation}
    p \left(i \right)=\kappa \delta \left(i \right) \ \ \ \textrm{if} \ \ \ i=0,\ \ \ p \left(i \right)=\frac{1}{a}\mathrm{e}^{-i/a} \ \ \ \textrm{if} \ \ \ i>0 ,
\end{equation}
where $\kappa$ describes the contribution from empty sky pixels relative to pixels from faint objects, and the wings of PSFs and galaxies. Given a sky background, $s$, and Gaussian noise, $\sigma$, the unnormalised observed flux ($I$) distribution is:
\begin{multline}
    p \left(I \right) \propto \frac{\kappa}{2 \pi {\sigma}^{2}} \textrm{exp}\left(-\frac{{\left(I-s \right)}^{2}}{2{\sigma}^2}\right) + \\ \frac{1}{a}\textrm{exp}\left(\frac{{\sigma}^2}{2{a}^2}\right)\textrm{exp}\left(-\frac{\left(I-s\right)}{a}\right) {\textrm{erf}}_{c}\left(\frac{\sigma}{a}-\frac{\left(I-s\right)}{\sigma}\right)
\end{multline}

The parameter space is explored using \textsc{emcee} \citep{foreman-mackey2013}, so a true sky background and its uncertainty are inferred. The flux for each star is then the sum of the background-subtracted 5.4 arcsecond (20 pixel) circular aperture, and the flux uncertainty includes Poisson noise, sky-background noise, and uncertainty in the subtracted background value. Finally we include a magnitude uncertainty of 0.003 in quadrature with the photometric uncertainty to account for possible systematic biases in bright stars, following \citet{burke2018}.

\item in each band, we simultaneously fit the zeropoints of all frames in that band. Since our goal is to measure variability, we can set one frame's zeropoint to that of the average value derived from the calibration stars' catalogued magnitudes \citep{desdr1}. For any combination of zeropoints, each calibration star's best-fit magnitude is found by minimising the ${\chi}^2$. Therefore, for $N$ epochs in a given band, we perform an optimisation procedure to infer $N-1$ zeropoints. Any stars with reduced ${\chi}^{2}$ above 100 are removed and the optimisation is repeated. This procedure is then iterated, removing stars with reduced ${\chi}^{2}$ above 50, 25, 10, 6, and 4. This preferentially removes the fainter stars in the sample, perhaps due to intrinsic variability, or because their photometry is more sensitive to seeing variations and contamination by nearby objects. Typically $\sim$80 per cent of stars are retained after these iterations, and the zeropoints are then sampled with \textsc{emcee}. Uncertainties on these zeropoints are $\sim$0.001-0.003 mag.

The relative differences between these zeropoints are consistent with the relative differences of DES pipeline zeropoints \citep{burke2018} to $<$0.01 mag. However, the absolute magnitudes can differ by up to 0.02 mag. This is expected since we attempted no aperture corrections on the star fluxes. Instead, we will apply the same 5.4 arcsecond apertures on our model PSF components, since we only want to constrain relative fluxes.

\item After fitting a nearby point source with a Moffat profile, the best seeing frame in each band is fit simultaneously with a combination of point sources and Sersic profiles. Astrometric registrations between the \textit{g}- and \textit{rizY}-bands are also modelled. Once the chains have converged, the best-fit model is used to set the alignment of all other frames (again with a nearby star fit for the frame's Moffat parameters), and finally the PSF and galaxy parameters are constrained from all frames simultaneously. These new best-fit parameters are used to constrain the best-fitting Moffat profiles and alignment on each frame individually, and all frames are again modelled simultaneously. Any poor-fitting frames are removed after visual inspection, and the Moffat fits and simultaneous-frame fits are repeated. The convergence of each chain is visually checked at each stage. This process provides a burnt-in chain of PSF positions and galaxy parameters.

\item For 50 samples of this model chain, the model PSF fluxes are determined in each frame through the same apertures as were applied to the zeropoint calibration stars (5.4 arcsecond circular apertures). However, since the Moffat parameters can affect these fluxes by $\sim$1\ per cent, we marginalise over all Moffat parameters and alignment offsets, by sampling these parameters at each step of the model chain. The concatenated chain of fluxes for PSFs and galaxies (the sum of the brightest 500 unconvolved pixels), provides the magnitudes and their uncertainties at each epoch. We add an ad hoc uncertainty of 0.005 mag in quadrature with the sampled uncertainty to account for any remaining systematics (poor centering of calibration stars within their apertures, lack of modelling of host galaxies, etc.), which brings the reduced ${\chi}^2$ for non-variable objects to unity.

\end{enumerate}

\subsection{Variability of DES systems}
\subsubsection{Removing stellar contaminants} \label{qsostarvar}
The most common contaminant in lens searches is quasars projected close to blue stars, mimicking doubly imaged quasars. While their optical colours can be similar, their SEDs vary towards redder wavelengths, as exploited by the modelling of unWISE pixels using \textit{Gaia} positions to remove such systems from lens searches \citep{lemon2019}. However, this modelling can only be applied to brighter systems with \textit{Gaia} detections and high signal-to-noise detections in \textit{WISE}.

\begin{figure}
	\includegraphics[width=\columnwidth]{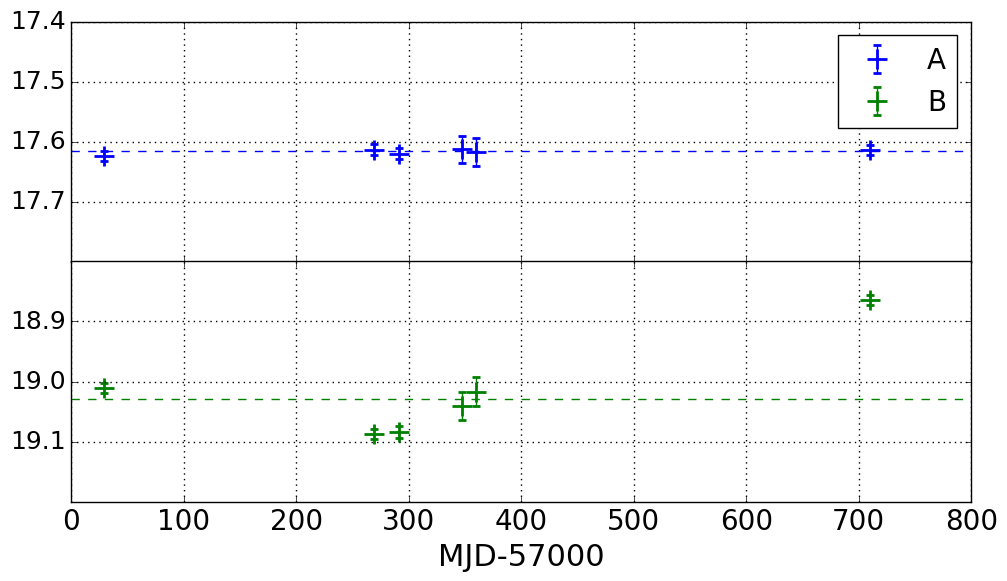}
    \caption{DES \textit{z}-band lightcurves for the projected star and quasar system DESJ0201-2043, clearly showing variability in the quasar component. Such lightcurves rely only on DES data, and as such are a useful way to remove contaminant systems from spectroscopic follow-up.}
    \label{fig:J0201z}
\end{figure}

\begin{figure}
	\includegraphics[width=\columnwidth]{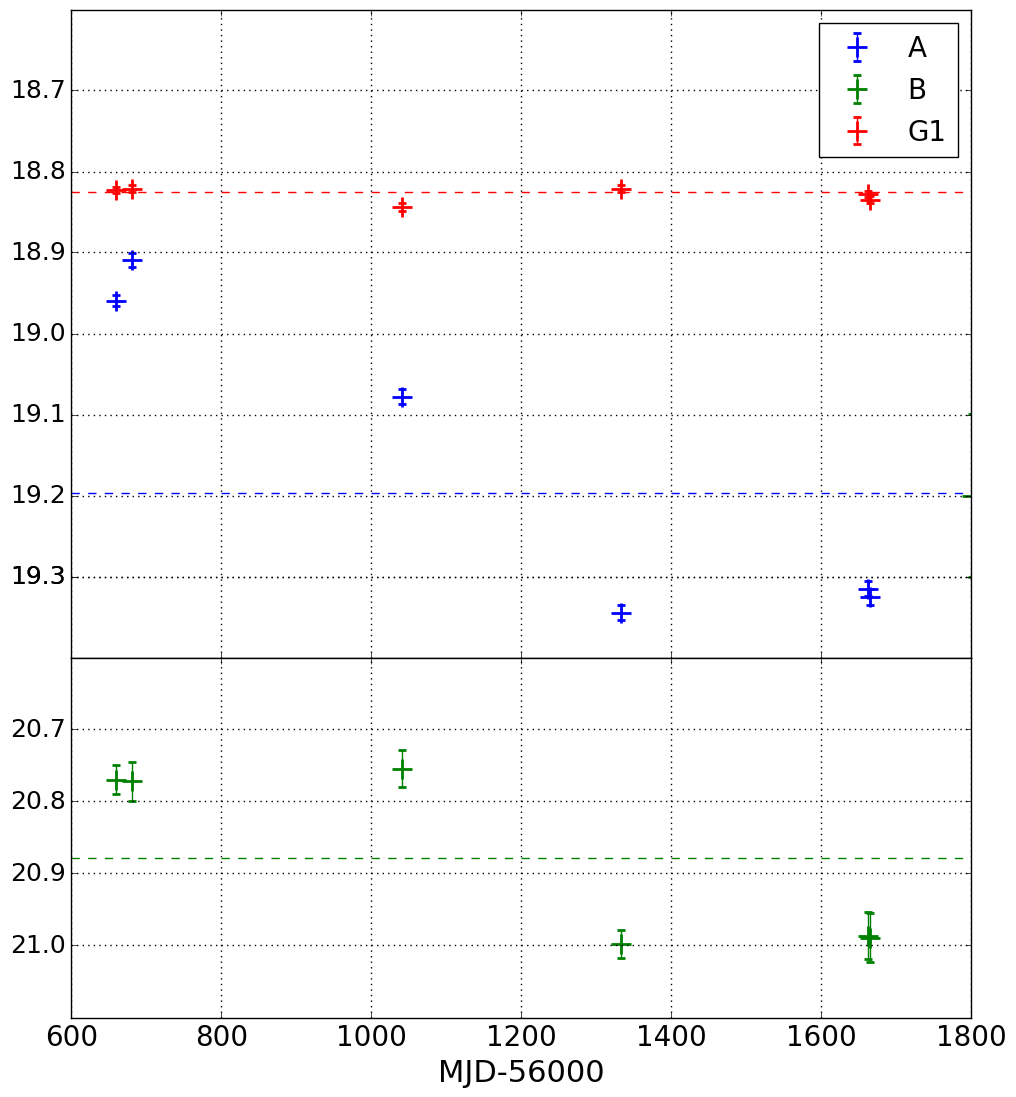}
    \caption{DES \textit{i}-band lightcurves for the lensed quasar system DESJ0150-4041, showing similar variations in the quasar images over a long baseline, and a lensing galaxy with consistent photometry.}
    \label{fig:J0150i}
\end{figure}

We investigate how variability information can be used to further remove contaminant systems including stars. We take a sample of 16 spectroscopically confirmed quasar and star pairs \citep[Appendix B,][]{strides1, anguita18} and derive their lightcurves following Section \ref{pipeline}. We also repeat the analysis for all known lensed quasars within the DES footprint\footnote{https://www.ast.cam.ac.uk/ioa/research/lensedquasars/}. Example lightcurves with representative sampling and photometric uncertainties are shown in Figures \ref{fig:J0201z} and \ref{fig:J0150i} for a quasar+star system and lensed quasar,  respectively. For our analysis we only retain systems with more than three photometric points over at least two observing seasons in at least three bands. To measure the variability of components in these systems, we take the reduced ${\chi}^2$ after fitting all photometry to the same magnitude. We expect values around 1 for non-variable components, but we also note that for magnitudes at the limiting single-epoch magnitude for the survey, the photometric errors naturally limit our detection of variability. Furthermore, there exist covariances between the photometric datapoints since the zeropoints were fit simultaneously, however we can safely neglect these since the zeropoints are much more well-determined than the relative photometry.

\begin{figure}
	\includegraphics[width=\columnwidth]{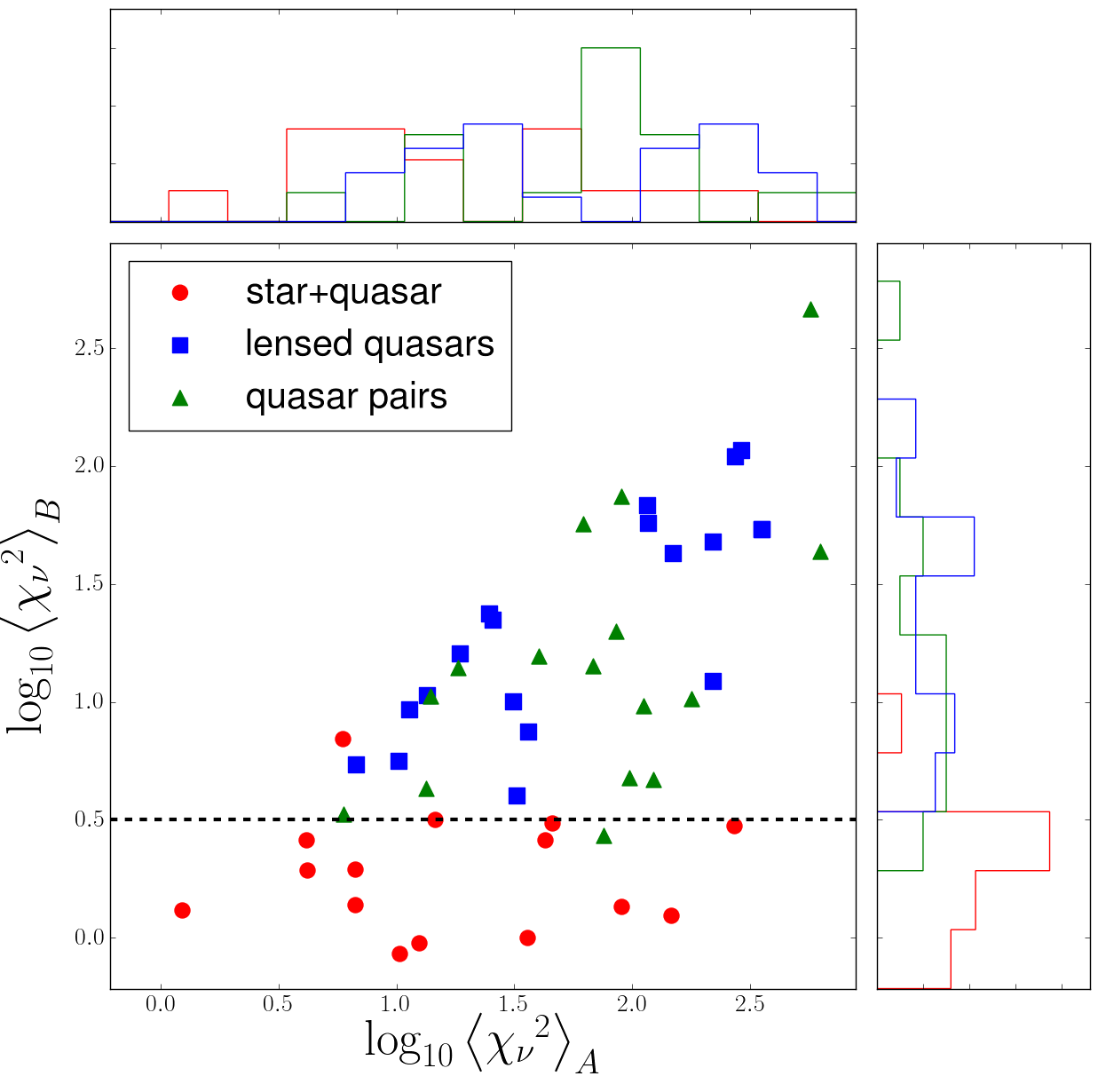}
    \caption{Average reduced ${\chi}^2$ over all bands containing three or more epochs of reliable photometry over at least two observing seasons. A cut of average $\textrm{log}_{10} \ {\chi}^2 >$0.5 retains all modelled lensed quasars, while removing 94 per cent of the contaminant systems containing stars. The more variable component (B) is plotted against the less variable component (A).} 
    \label{fig:chisquaredvar}
\end{figure}

Figure \ref{fig:chisquaredvar} shows the average reduced ${\chi}^{2}$ across all measured bands for the two most variable point sources in a system, for a sample of 15 quasar and star systems, 16 lensed quasars, and 16 quasar pairs. The quasar-pair sample may contain lensed quasars with faint, yet currently undetected, lensing galaxies, and are discussed further in Section \ref{niqs}. We expect the variability of the less variable object to be a good indicator of whether the system has only quasar components, or a non-variable quasar component. A cut retaining systems with average reduced ${\chi}^{2} > 3.16$ removes 15 of the 16 stellar contaminant systems, while retaining all lensed quasars. Assuming that we can extend this analysis to all 24 systems we have spectroscopically classified as containing stars with the DES 6 year data (but are not currently able to, due to lacking the number of good-fitting epochs in at least three bands), we would expect that the suggested cut can remove 22 systems. This would reduce our contaminant systems to 13, composed of 2 systems with stars, and the rest of star-forming galaxies and quasar-galaxy projections (i.e., at different redshifts without strong lensing). Our confirmation rate of lensed quasars and quasar pairs would then increase from 34-45 per cent to 51-70 per cent, with the spread due to the inconclusive systems. Further investigation of colour-based variability may allow for separation of variable stars and quasars.

The top histogram of Figure \ref{fig:chisquaredvar} shows the variability measure for the most variable component in each system. For the quasar and star pairs this is always the quasar component, and for the lensed quasars, this is generally the brighter image. The unlensed quasars from the quasar+star pairs are less variable than their lensed counterparts. This is not due to the former set having fainter apparent magnitudes and hence being less significantly variable through the ${\chi}^2$ statistic, as both sets have similar distributions of observed brightnesses. It is more likely explained by a combination of three factors: (i) the magnification due to the lensing implies that intrinsic magnitudes of lensed quasars are fainter than the counterparts in the quasar+star systems, and less luminous quasars are well-known to vary more \citep[e.g.,][]{hook1994, szymon2016, li2018}; (ii) lensed quasars are susceptible to additional extrinsic variations through microlensing; and (iii) our unlensed quasar sample has a lower redshift distribution than the lensed quasars. Given that bluer quasar emission is intrinsically more variable, for a given observed wavelength range, a higher-redshift quasar will be more variable. Furthermore, there is a weak trend of increasing variability with increasing redshift at a fixed rest-frame wavelength \citep{li2018}.



\subsubsection{J0235-2433}
\begin{figure}
	\includegraphics[width=\columnwidth]{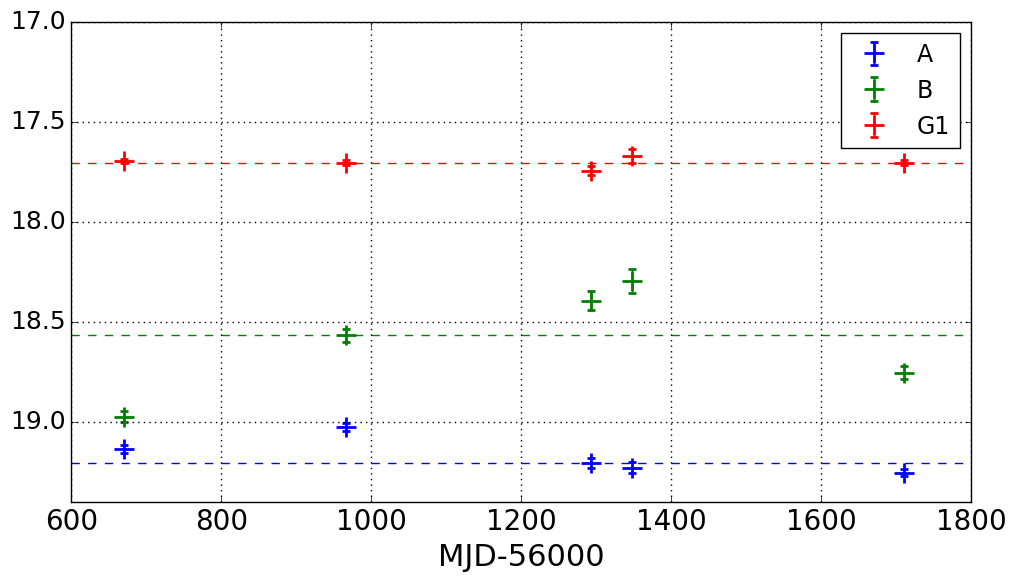}
    \caption{DES \textit{Y}-band lightcurves for the lensed quasar system J0235-2433, showing a likely microlensing event in image B over 3 years. Similar variations are seen in the other bands, and cannot be explained by the time delay difference without invoking multiple large variation and short-lived flares in the source quasar.}
    \label{fig:J0235}
\end{figure}

This doubly imaged lensed quasar was suspected to have an image undergoing a long-term microlensing event due to a large discrepancy between the image flux ratio in the DES and Pan-STARRS data \citep{lemon2018}. The DES lightcurves corroborate this interpretation. In all bands, image A decreases in brightness by $\sim$0.3 mag from Year 2 to Year 4, while image B increases over the first year of this same period by 0.3 mag and drops by 0.6 mag in the following year. The \textit{z}- and \textit{Y}-band lightcurves include an epoch in Year 1, showing the two images are of similar magnitude, while in Year 3 the difference is 0.9 mag. If the time delay were comparable to the baseline of our observations, then this could simply be an effect caused by the images sampling the source quasar at entirely different epochs. However, the source redshift is relatively low ($z=$1.44), the lens galaxy is particularly bright implying a low redshift, and the separation is modest (2.04 arcseconds). Simple lens models for such systems are expected to have time delays under 50 days, much shorter than the 3 year observation baseline. The \textit{Y}-band lightcurves (the band with the longest baseline) are shown in Figure \ref{fig:J0235}, clearly showing that image B has undergone variations inconsistent with that of image A. 

\subsubsection{Nearly Identical Quasar Pairs} \label{niqs}
The \citet{om10} mock catalogue for lensed quasars with bright images, i.e. detected by \textit{Gaia}, predicts $\sim$17 per cent of lensing galaxies will have $i$-band magnitudes fainter than 22.4, namely a magnitude brighter than the DES data release 1 coadd magnitude limit ($MAG\_APER\_4$, 1.95 arcsecond diameter, S/N=10). We might expect then that some of the quasar pairs identified in this work, and previous STRIDES publications, are lensed quasars. Obtaining deep, high-resolution imaging of all such pairs would be an expensive, and potentially inefficient, project. However, an indicator of the gravitational lensing hypothesis would be similar long-term variability in the multiple components of a system. Our variability pipeline allows us to consider such objects over a baseline of 4 years. 

We model all the newly discovered quasar pairs where both components are at the same redshift (see Table \ref{tab:qsopairs}), including those from \citet{anguita18}. The variability metric described in Section \ref{qsostarvar} is plotted for the components of these NIQs in Figure \ref{fig:chisquaredvar}. They are clearly detected as having multiple variable components, and we might expect that if such systems were lensed quasars, the variability of the images would be similar. However, this neglects the effects of extrinsic variability due to microlensing, difference in time sampling of the true quasar lightcurve due to the time delay, and the different photometric precision due to flux differences. We inspect each pair looking for similar long-term variability, taking into account a possible time delay causing a shift in the variability. Candidates showing similar variability are DESJ0229+0320 and DESJ2215-5204, and those showing seemingly uncorrelated variability are DESJ0122+0358 and DESJ0313-2546. These latter two systems have SOAR $z$-band imaging, which show no signs of lensing galaxies after PSF subtraction. Therefore the former systems should be given priority for high-resolution imaging follow-up. We warn that extrinsic variability, sampling differences due to the time delay, and chance correlations in distinct quasars, could confuse the signs of lensing.

\subsection{Variability Selection Bias}
Our lightcurves have currently only been used to demonstrate the proof of concept of using variability to efficiently discover lensed quasars. Applying such a routine to all close pairs in DES is unfeasible due to the computation time of our pipeline. \citet{kochanek2006} suggested difference imaging would be an efficient way to select lensed quasars as extended variable objects. We expect difference imaging will become an effective way to find lensed quasars in current and upcoming full-sky surveys. This naturally raises the question of selection effects from selecting the most variable systems. 

\begin{figure}
	\includegraphics[width=\columnwidth]{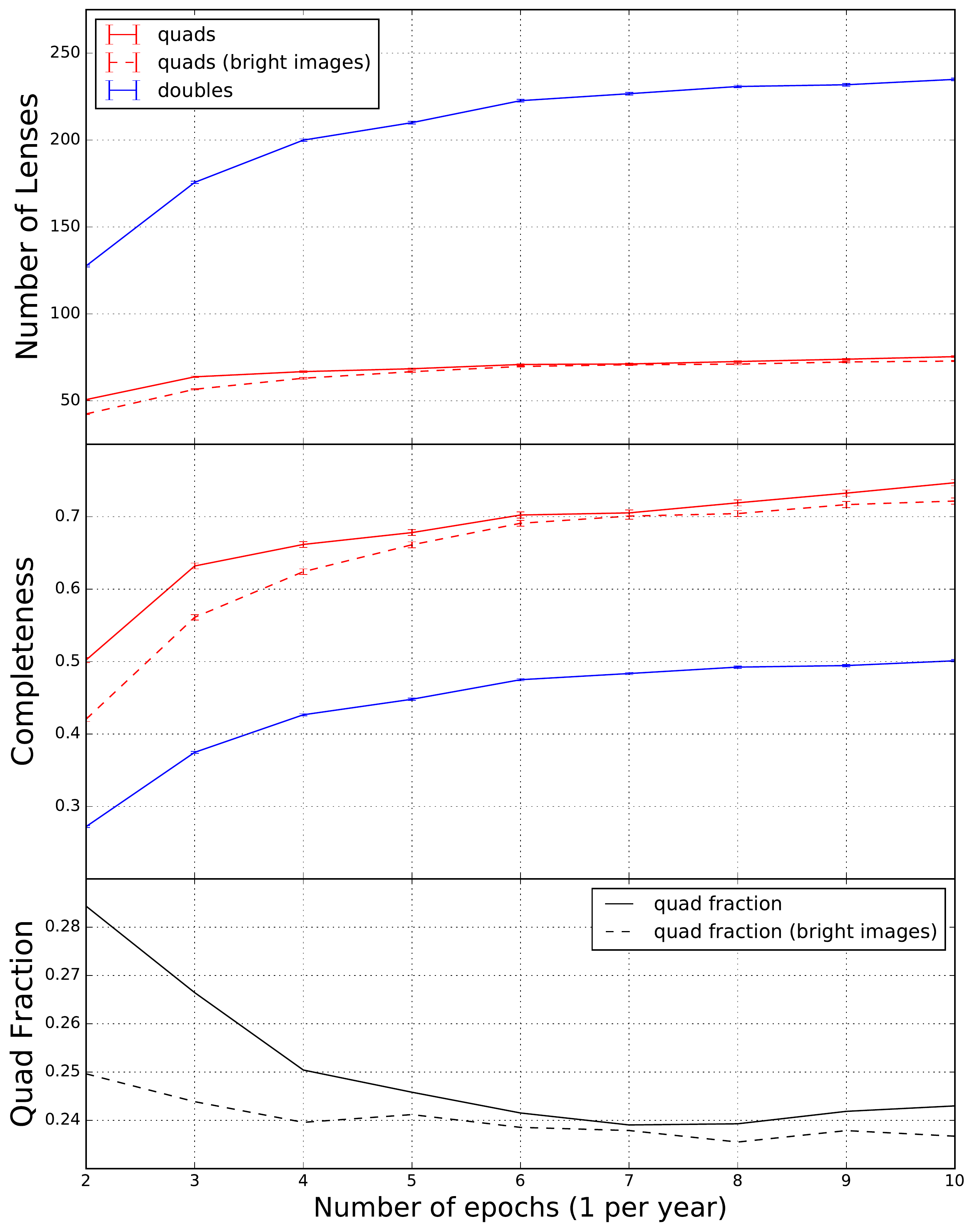}
    \caption[Predicted numbers of quads and doubles to be discovered through variability from DES $i$-band multi-epoch imaging]{\textit{Top}: number of doubles and quads in 5,000 square degrees passing the variability threshold condition for at least two images as a function of the number of epochs each separated by exactly one year. \textit{Middle}: fraction of systems compared to all systems with at least two images brighter than the measured single-epoch $i$-band DES depth of 22.78. \textit{Bottom}: quad fraction of lenses expected to be discovered with variability techniques. This is much higher than the magnitude-limited sample of lenses, which would have a quad fraction of $\sim$ 17.7\%. Dashed lines for quads are limiting variability detection to just the two brightest images. This reduces the quad fraction by <1\% for the DES 6 year survey, implying the magnification bias alone causes a 6\% increase in quad fraction over the magnitude-limited sample.}
    \label{fig:varselection}
\end{figure}

Systems with the faintest intrinsic sources will have the most variable images. For a magnitude-limited survey there is a trade-off between faint sources being more variable, but also having less precise photometry. In practice, therefore, we expect quasar images with brightnesses near the magnitude limit of a survey to only be robustly detected through their variability if they are high-magnification images. This will favour the selection of quads. We explore such a bias using mock lightcurves for a DES-like survey, and determine how the reduced ${\chi}^{2}$ cut from our DES analysis affects the discovered population. We take the OM10 mocks and generate lightcurves using the Damped Random Walk parameters and uncertainties from \citet{macleod2010} based on source magnitude and redshift. The photometric uncertainties are estimated based on all our $i$-band measured uncertainties for lensed quasar $i$-band image magnitudes between 18 and 22.5. Typical uncertainties for images with $i$=19, 20, 21, 22, and 23 are 0.01, 0.02, 0.04, 0.15, and 0.77 respectively. We assume a pessimistic cadence of one observation per year to understand the selection effects from mock DES data. The lightcurves are shifted by the relevant time delays and single-epoch magnitudes are sampled. The ${\chi}^{2}$ statistic is generated as in Section \ref{qsostarvar} and a system is counted as discovered if at least two images have ${\chi}^{2}$>3.16. In Figure \ref{fig:varselection}, the number of quads and doubles passing this criterion is shown after different numbers of epochs. The numbers naturally increase with more epochs. For a 6 year survey like DES, we would expect variability to discover $\sim$70 quads, and $\sim$220 doubles. The middle panel of Figure \ref{fig:varselection} shows these numbers as a fraction of the total number of quads and doubles that have at least two images brighter than the measured 10$\sigma$ single-epoch $i$-band depth of DES, i.e. 22.78. As expected, the quad fraction is higher due to magnification bias. Requiring just the two brightest images of quads to be the ones detected for variability does not significantly change the results (as shown by the dashed lines in Figure \ref{fig:varselection}), so the higher quad fraction is not due to having more images in quads, and hence having a higher probability of detecting at least two variable images.

The absolute numbers are conservative since we have only used one band. Using the $g$-band would be best for detecting variability, however this is the band for which we have found the PSF most difficult to model. Furthermore, we have only considered intrinsic variations as a method for detecting lensed quasars. Extrinsic variations from microlensing will make these lightcurves more variable, making our estimates more conservative.

This variability bias will have an impact on the subset of lensed quasars that will have their time delays measured via LSST. Early predictions from OM10 suggested up to 3,000 lensed quasars will have measured time delays in LSST. A tested prediction came from the Time Delay Challenge, which created thousands of mock LSST lensed quasar lightcurves with known time delays, then given to the community for blind time delay retrieval \citep{dobler2015, liao2015}. Based on teams' correct retrievals of time delays, approximately 400 robust measurements will be expected from LSST; however, the mock lightcurves drew variability parameters uniformly from observed ranges, rather than using their correlations with intrinsic luminosity. It is unclear how this would affect the numbers of robust time-delay measurements, though the quad fraction within this sample will be higher because of the same causes of variability selection bias. A multiplicity-biased sample of lensed quasars with time delays, as expected from LSST, can lead to biases in cosmological parameters without careful consideration of priors in lens modelling \citep{collett2016}.

\section{Conclusions} \label{conclusion}
We have presented the spectroscopic and high-resolution imaging follow-up from the STRIDES 2017-2018 campaign, discovering 10 new lensed quasars and 12 quasar pairs. Ten of these pairs have quasars at the same redshift, and further imaging might reveal lensing galaxies. Of particular interest in the lensed quasar sample is a quadruply imaged quasar, DESJ0053-2012, with a source redshift of 3.8, and a triply imaged 6.8 arcsecond separation lensed quasar with all images detected by \textit{Gaia}.

After the end of the follow-up campaign, we applied a parametric modelling pipeline (any combination of point sources and galaxies) to determine the level of variability in individual components in any system. Applying this to known lensed quasars and spectroscopically confirmed quasar+star pairs in the DES footprint, we provide a prescription for removing 94 per cent of the latter from future spectroscopic follow-up campaigns, while retaining all lensed quasars. However, we note that future campaigns targeting a complete sample of lensed quasars with difference imaging techniques---as might be applied to the LSST---will need to take into account the variability bias of their search. This will require a good understanding of the variability distribution of the sources, and the magnifications due to realistic lensing galaxies. Such searches will be biased towards high-magnification systems, since there exists a strong anti-correlation between intrinsic luminosity and variability. 

In our 2016/2017 campaign our confirmation rate was 6-35 per cent (with spread due to NIQs and inconclusive systems), while in 2017/2018 it was (10-27)/65=15-41 per cent. Future selection should show significant improvement in efficiency based on our variability modelling technique and new auxiliary data sets, such as \textit{Gaia} data release 2. The Dark Energy Survey recently completed survey operations after six years so we will soon extend the baseline of these studies by 50 per cent, which will improve the certainty of contaminant systems, prioritisation of follow-up of quasar pairs as potential lensed quasars, and detection of lensing galaxies through deeper stacked imaging. Future improvements on extracting variability from single-epoch images may come from non-parametric reconstructions of the PSF \citep{chen2016} and/or extended light emission \citep{cantale2016}, or by modelling difference images \citep{kochanek2006, lacki2009, chao2019}.

\section*{Acknowledgements}

This work has made use of data from the European Space Agency (ESA) mission {\it Gaia} (\url{https://www.cosmos.esa.int/gaia}), processed by the {\it Gaia} Data Processing and Analysis Consortium (DPAC, \url{https://www.cosmos.esa.int/web/gaia/dpac/consortium}). Funding for the DPAC has been provided by national institutions, in particular the institutions participating in the {\it Gaia} Multilateral Agreement. Some of the data presented herein were obtained at the W. M. Keck Observatory, which is operated as a scientific partnership among the California Institute of Technology, the University of California and the National Aeronautics and Space Administration. The Observatory was made possible by the generous financial support of the W. M. Keck Foundation. 

Funding for the DES Projects has been provided by the U.S. Department of Energy, the U.S. National Science Foundation, the Ministry of Science and Education of Spain, the Science and Technology Facilities Council of the United Kingdom, the Higher Education Funding Council for England, the National Center for Supercomputing Applications at the University of Illinois at Urbana-Champaign, the Kavli Institute of Cosmological Physics at the University of Chicago, the Center for Cosmology and Astro-Particle Physics at the Ohio State University, the Mitchell Institute for Fundamental Physics and Astronomy at Texas A\&M University, Financiadora de Estudos e Projetos, Funda{\c c}{\~a}o Carlos Chagas Filho de Amparo {\`a} Pesquisa do Estado do Rio de Janeiro, Conselho Nacional de Desenvolvimento Cient{\'i}fico e Tecnol{\'o}gico and the Minist{\'e}rio da Ci{\^e}ncia, Tecnologia e Inova{\c c}{\~a}o, the Deutsche Forschungsgemeinschaft and the Collaborating Institutions in the Dark Energy Survey. 

The Collaborating Institutions are Argonne National Laboratory, the University of California at Santa Cruz, the University of Cambridge, Centro de Investigaciones Energ{\'e}ticas, 
Medioambientales y Tecnol{\'o}gicas-Madrid, the University of Chicago, University College London, the DES-Brazil Consortium, the University of Edinburgh, the Eidgen{\"o}ssische Technische Hochschule (ETH) Z{\"u}rich, Fermi National Accelerator Laboratory, the University of Illinois at Urbana-Champaign, the Institut de Ci{\`e}ncies de l'Espai (IEEC/CSIC), the Institut de F{\'i}sica d'Altes Energies, Lawrence Berkeley National Laboratory, the Ludwig-Maximilians Universit{\"a}t M{\"u}nchen and the associated Excellence Cluster Universe, the University of Michigan, the National Optical Astronomy Observatory, the University of Nottingham, The Ohio State University, the University of Pennsylvania, the University of Portsmouth, SLAC National Accelerator Laboratory, Stanford University, the University of Sussex, Texas A\&M University, and the OzDES Membership Consortium.

Based on observations obtained at the Southern Astrophysical Research (SOAR) telescope, which is a joint project of the Minist\'{e}rio da Ci\^{e}ncia, Tecnologia, Inova\c{c}\~{o}es e Comunica\c{c}\~{o}es (MCTIC) do Brasil, the U.S. National Optical Astronomy Observatory (NOAO), the University of North Carolina at Chapel Hill (UNC), and Michigan State University (MSU).
V.M. acknowledges the support of the Centro de Astrof\'{\i}sica de Valpara\'{\i}so.

The DES data management system is supported by the National Science Foundation under Grant Numbers AST-1138766 and AST-1536171.
The DES participants from Spanish institutions are partially supported by MINECO under grants AYA2015-71825, ESP2015-66861, FPA2015-68048, SEV-2016-0588, SEV-2016-0597, and MDM-2015-0509, 
some of which include ERDF funds from the European Union. IFAE is partially funded by the CERCA program of the Generalitat de Catalunya. Research leading to these results has received funding from the European Research Council under the European Union's Seventh Framework Program (FP7/2007-2013) including ERC grant agreements 240672, 291329, and 306478. We  acknowledge support from the Brazilian Instituto Nacional de Ci\^enciae Tecnologia (INCT) e-Universe (CNPq grant 465376/2014-2).

Based in part on observations at Cerro Tololo Inter-American Observatory, National Optical Astronomy Observatory, which is operated by the Association of Universities for Research in Astronomy (AURA) under a cooperative agreement with the National Science Foundation. 

This work is supported by the Swiss National Science Foundation (SNSF). This project has received funding from the European Research Council (ERC) under the European Union's Horizon 2020 research and innovation program (COSMICLENS: grant agreement No 787886). AA was supported by a grant from VILLUM FONDEN (project number 16599). This project is partially funded by the Danish council for independent research under the project ``Fundamentals of Dark Matter Structures'', DFF--6108-00470. TA acknowledges support from Proyecto FONDECYT N: 1190335. AJS acknowledges support from the National Aeronautics and Space Administration through the Space Telescope Science Institute grant HST-GO-15320 and from University of California, Los Angeles graduate division through a dissertation year fellowship. CDF and GCFC acknowledge support for this work from the National Science Foundation under Grant No. AST-1715611. TT acknowledges support by NSF through grants  AST-1450141 and AST-1906976 and by the Packard Foundation through a Packard Fellowship. 

The authors wish to recognize and acknowledge the very significant cultural role and reverence that the summit of Maunakea has always had within the indigenous Hawaiian community.  We are most fortunate to have the opportunity to conduct observations from this mountain.

This manuscript has been authored by Fermi Research Alliance, LLC under Contract No. DE-AC02-07CH11359 with the U.S. Department of Energy, Office of Science, Office of High Energy Physics.




\bibliographystyle{mnras}
\bibliography{papers} 




\appendix
\section{All observed targets}
Tables \ref{tab:inconc} and \ref{tab:contaminants} list all inconclusive systems and contaminant systems, respectively, each with spectroscopic follow-up. While all systems show multiple components in the imaging data, spectroscopy does not always resolve these components and are marked as blended in the outcome. Systems are inconclusive either due to blending or due to lack of signal-to-noise.

\begin{table*}
	\centering
	\caption{Inconclusive candidates. Selection: G1: \textit{Gaia} 1, G2: \textit{Gaia} 2, V: Variability, C: component fitting.}
	\label{tab:inconc}
	\begin{tabular}{lcccccl} 
		\hline
		Name & R.A. (J2000) & Dec. (J2000) & Selection & spectrum & imaging & outcome\\
		\hline
        DESJ0149-6532 & 27.2900 & -65.54034 & G1 & EFOSC2 & - & low flux, inconclusive \\
        DESJ0402-3237 & 60.5626 & -32.6260 & V & EFOSC2 & SOAR & $z=$1.28 QSO blended (2 PSFs in SOAR data)\\
        DESJ0402-4220 & 60.5922 & -42.3482 & V & EFOSC2 & SOAR & $z=$2.88 QSO blended, (PSF + galaxy in SOAR)\\
        DESJ0428-2933 & 67.1001 & -29.5552 & G1 & EFOSC2 & - & insecure $z=$0.74 and inconclusive QSO companion\\
        DESJ0451-2147 & 72.80567 & -21.7967 & G1 & EFOSC2 & - & possible NIQ $z=$1.07? inconclusive\\
        DESJ0508-2748 & 77.0133 & -27.8053 & G2 &  EFOSC2/ESI & SOAR/NIRC2 & $z=$1.14 blended (2 PSFs in NIRC2 data)\\
        DESJ0551-4629 & 87.9493	& -46.4982 & V & EFOSC2 & SOAR & blended $z=$1.21 (insecure) QSO  (2 PSFs in SOAR data) \\
		\hline
	\end{tabular}
\end{table*}

\begin{table*}
	\centering
	\caption{Spectroscopically confirmed contaminant systems. Selection abbreviation is as in Table 1.}
	\label{tab:contaminants}
	\begin{tabular}{lcccccl} 
		\hline
		Name & R.A. (J2000) & Dec. (J2000) & Selection & spectrum & imaging & outcome\\
		\hline
        DESJ0007+0053 & 1.7917 & 0.8914 & V & ESI & - & $z=$0.32 QSO blended  \\
        DESJ0050-0432 & 12.5106 & -4.5454 & V & ESI & - & two galaxies/low flux \\
        DESJ0058-3947 & 14.55486 & -39.7899 & G1 & EFOSC2 & SOAR & $z=$0.51 QSO + star \\
        DESJ0100-3406 & 15.1283 & -34.1092 & C &  EFOSC2 & SOAR & $z=$1.73 QSO blended spectrum\\
        DESJ0140-2234 & 25.1347 & -22.5829 & V & ESI & - & $z=$0.559 galaxies \\
        DESJ0140-1700 & 25.2159 & -17.0004 & V & EFOSC2 & - & $z=$0.294 galaxies\\
        DESJ0201-2043 & 30.3689 & -20.7192 & C, V & EFOSC2 & SOAR & $z=$0.65 QSO + star\\
        DESJ0210-4612 & 32.7282 & -46.2144 & G2 &  EFOSC2 & SOAR & $z=$1.81 QSO + other\\
        DESJ0214-2146 & 33.5046 & -21.7733 & C &  EFOSC2 & - & resolved $z=$0.84 QSO + $z=$0.84 galaxy component\\
        DESJ0223-3312 & 35.8888	& -33.2103 & C &  EFOSC2 & SOAR & $z=$0.29 emission line galaxies \\
        DESJ0243-2410 & 40.8640 & -24.1720 & V & EFOSC2 & - & $z=$0.08 galaxies \\
        DESJ0248-0412 & 42.0517 & -4.2067 & V & ESI & - & $z=$0.239 galaxies\\
        DESJ0248-2632 & 42.1457 & -26.5438 & V & EFOSC2 & SOAR & blended QSO (broad emission at 5350\AA)\\
        DESJ0303-5055 & 45.9475 & -50.9177 & V & EFOSC2 & SOAR & star + other\\
        DESJ0316-5228 & 49.1702	& -52.4802 & V & EFOSC2 & SOAR & $z=$0.89 QSO + star\\
        DESJ0343-3309 & 55.9238 & -33.1556 & G1, G2 &  EFOSC2 & SOAR & $z=$1.58 QSO + star\\
        DESJ0350-4014 & 57.5075 & -40.2489 & C &  EFOSC2 & - & $z=$0.88 QSO + star\\
        DESJ0354-1944 & 58.6742	& -19.7469 & C &  EFOSC2 & - & $z=$0.244 galaxies \\
        DESJ0418-5722 & 64.7172 & -57.3698 & G1 & EFOSC2 & - & $z=$2.02 QSO + other \\
        DESJ0448-2719 & 72.1800 & -27.3255 & G2 &  ESI & SOAR & $z=$0.57 QSO + star\\
        DESJ0455-5412 & 73.9222 & -54.2067 & G1 & EFOSC2 & SOAR & QSO ($z=$1.01?) + other\\
        DESJ0456-2823 & 74.1955 & -28.3910 & V & EFOSC2 & SOAR & $z=$1.44 QSO + star\\
        DESJ0457-4748 & 74.2835 & -47.8164 & V & EFOSC2 & - & $z=$0.32 galaxies \\
        DESJ0511-5334 & 77.9206 & -53.5824 & V & EFOSC2 & - & $z=$1.55 QSO + star\\
        DESJ0512-1817 & 78.1131 & -18.2983 & G1 & ESI & - & $z=$0.343 emission line galaxies \\
        DESJ0520-4437 & 80.0602	& -44.6292 & V & EFOSC2 & SOAR & $z=$0.345 QSO + other\\
        DESJ0531-6012 & 82.9947	& -60.2092 & V & EFOSC2 & - & $z=$0.29 QSO + star\\
        DESJ0542-4911 & 85.6279	& -49.1879 & V & EFOSC2 & - & $z=$1.20? QSO + star\\
        DESJ0559-3428 & 89.8474 & -34.4721 & G1 & EFOSC2 & SOAR & $z=$0.75 QSO + other \\
        DESJ2047-4801 & 311.8621 & -48.0299 & G1 & EFOSC2 & - & $z=$0.71 QSO + star\\
        DESJ2104-4228 & 316.1840 & -42.4815 & G2 &  EFOSC2 & - & $z=$1.58 QSO + star\\
        DESJ2120-4239 & 320.1450 & -42.6514 & C &  EFOSC2 & - & $z=$0.83 QSO + star\\
        DESJ2139-4331 & 324.9881 & -43.5172 & G2 &  EFOSC2 & - & $z=$0.48 QSO + star\\
        DESJ2210-4536 & 332.7210 & -45.6084 & V & EFOSC2 &  - & $z=$0.82 QSO + star\\
        DESJ2248-4903 & 342.1560 & -49.0530 & V & EFOSC2 & - & $z=$0.55 QSO + star\\
        DESJ2332-4934 & 353.0430 & -49.5685 & G1, G2 &  EFOSC2 & - & $z=$0.74 QSO + star\\
		\hline
	\end{tabular}
\end{table*}

\section{Astrometry and Photometry}
Table \ref{tab:astrophotometry} provides the astrometry from the best available imaging data and DES photometry for each lens system. The uncertainties on quasar photometry reflect the standard deviation of multi-epoch magnitude values, or in the case of only one epoch fitting being retained, the uncertainty on that single-epoch magnitude. In the case of the lens galaxy, the photometry and uncertainty are given for the most precise single-epoch band.

\begin{table*}
  \centering
  \caption{Astrometry and photometry of all confirmed lensed quasars. Magnitudes are in the AB sytem. All photometry is from DES. Astrometry is from NIRC2 (DESJ0246, DESJ0245, DESJ0340), SOAR (J0053, J0150, J0407, J0501, J0600), or DES (DESJ0112). The photometric uncertainties are the standard deviation of magnitudes across all measured epochs. Stated flux ratios are from the best imaging data. }
  \label{tab:astrophotometry}
  \begin{tabular}{cccccccccc}
  \hline
  & component & $\alpha$ ($\arcsec$) & $\delta$ ($\arcsec$) & flux ratio & $g$ & $r$ & $i$ & $z$ & $Y$\\
  \hline

J0053-2012 & A &  1.828 $\pm$  0.001  &  -0.662  $\pm$  0.001  &  10.24 &20.22$\pm$0.04 & 19.44$\pm$0.02 & 19.32$\pm$0.03 & 19.06$\pm$0.02 & 19.16$\pm$0.01 \\
& B &  -0.263$\pm$0.001  &  -0.264$\pm$0.001   &  10.56 & 20.12$\pm$0.04 & 19.34$\pm$0.01 & 19.23$\pm$0.03 & 18.99$\pm$0.02 & 19.06$\pm$0.01 \\
& C &  -0.981$\pm$0.001  &  0.472$\pm$0.001  &  9.61 & 20.30$\pm$0.06 & 19.49$\pm$0.01 & 19.37$\pm$0.03 & 19.12$\pm$0.03 & 19.23$\pm$0.01 \\
& D &  0.95$\pm$0.01  &  1.46$\pm$0.01  &  1.0 & 22.49$\pm$0.07 & 21.92$\pm$0.04 & 21.76$\pm$0.05 & 21.63$\pm$0.14 & 21.29$\pm$0.12 \\
& G1 &  0.86$\pm$0.02  &  0.90$\pm$0.03  & ---  & 23.53$\pm$0.37 & 21.65$\pm$0.06 & 21.22$\pm$0.05 & 20.25$\pm$0.05 & 20.12$\pm$0.09 \\
& G2 &  -2.39$\pm$0.02  &  -1.91$\pm$0.02  & --- & 22.58$\pm$0.04 & 21.80$\pm$0.03 & 20.98$\pm$0.03 & 20.44$\pm$0.03 & 20.28$\pm$0.03 \\
\hline
J0112-1650 & A &  0.928$\pm$0.001  &  0.140$\pm$0.001  &  2.94 & 20.64$\pm$0.03 & 20.10$\pm$0.05 & 20.08$\pm$0.04 & 19.84$\pm$0.16 & 19.67$\pm$0.27 \\
& B &  -0.325  $\pm$  0.002  &  -0.486  $\pm$  0.002   &  1.74 & 20.99$\pm$0.13 & 20.70$\pm$0.06 & 20.65$\pm$0.09 & 20.55$\pm$0.08 & 20.59$\pm$0.75 \\
& C &  -0.089  $\pm$  0.004  &  0.684  $\pm$  0.004  &  1.0 & 23.39$\pm$0.20 & 21.78$\pm$0.09 & 21.25$\pm$0.02 & 20.96$\pm$0.06 & 21.05$\pm$0.36 \\
& G1 &  0.00  $\pm$  0.01  &  0.00  $\pm$  0.01  & --- & --- & 20.55$\pm$0.05 & 19.78$\pm$0.04 & 19.24$\pm$0.04 & 18.52$\pm$0.10 \\
\hline
J0150-4041 & A &  -1.643$\pm$0.001 & -0.003$\pm$0.001  &  4.74 & 19.78$\pm$0.04 & 19.64$\pm$0.13 & 19.19$\pm$0.18 & 19.01$\pm$0.16 & 18.98$\pm$0.01 \\
 & B &  1.163$\pm$0.005  &  0.028$\pm$0.005   &  1.0 & 21.31$\pm$0.01 & 21.20$\pm$0.09 & 20.87$\pm$0.11 & 20.77$\pm$0.13 & 20.78$\pm$0.02 \\
 & G1 &  0.480$\pm$0.004  &  -0.025$\pm$0.003  & --- & 21.38$\pm$0.02 & 19.43$\pm$0.01 & 18.82$\pm$0.01 & 18.43$\pm$0.01 & 18.37$\pm$0.01 \\
\hline
J0245-0556 & A & 0.00 & 0.00 & 1.46 & 19.38$\pm$4.89 & 19.24$\pm$0.04 & 19.27$\pm$0.05 & 19.42$\pm$0.05 & 19.44$\pm$0.04 \\
           & B & 1.1300$\pm$0.005 & -1.5300$\pm$0.005 & 1.0 & 19.97$\pm$0.01 & 19.73$\pm$0.06 & 19.80$\pm$0.13 & 19.93$\pm$0.08 & 20.00$\pm$0.05 \\
           & G & 0.9148$\pm$0.005 & -1.196$\pm$0.005 & - & 20.99$\pm$0.03 & 19.57$\pm$0.01 & 18.92$\pm$0.01 & 18.69$\pm$0.01 & 18.55$\pm$0.02 \\
\hline
J0246-1845 & A & 0.00 & 0.00 & 3.34 & 18.66$\pm$0.01 & 18.63$\pm$0.01 & 18.39$\pm$0.01 & 18.49$\pm$0.03 & 18.53$\pm$0.02 \\
           & B & 0.0430$\pm$0.005 & -0.9966$\pm$0.005 & 1.00 & 19.38$\pm$0.01 & 19.39$\pm$0.01 & 19.11$\pm$0.01 & 19.15$\pm$0.01 & 19.20$\pm$0.04 \\
           & G & 0.011$\pm$0.005 & -0.709$\pm$0.005 & --- & --- & --- & --- & --- & ---\\
\hline
J0340-2545 & A & 0.00 & 0.00 & 9.59 & 18.42$\pm$0.06 & 18.44$\pm$0.02 & 18.31$\pm$0.07 & 18.49$\pm$0.08 & 18.66$\pm$0.02 \\
& B & -3.102$\pm$0.005 & 5.985$\pm$0.005 & 3.49 & 19.85$\pm$0.13 & 19.74$\pm$0.07 & 19.51$\pm$0.06 & 19.82$\pm$0.08 & 19.84$\pm$0.11 \\
& C & 0.807$\pm$0.005 & 3.078$\pm$0.005 & 1.0 & 21.12$\pm$0.09 & 21.18$\pm$0.06 & 21.01$\pm$0.06 & 21.42$\pm$0.25 & 21.68$\pm$0.49 \\
& G1 & 0.720$\pm$0.005 & 2.758$\pm$0.005 & --- & 22.51$\pm$0.10 & 20.46$\pm$0.01 & 19.67$\pm$0.01 & 19.24$\pm$0.01 & 19.08$\pm$0.01 \\
& G2 & -2.960$\pm$0.005 & 3.263$\pm$0.005 & --- & 21.91$\pm$0.01 & 20.23$\pm$0.01 & 19.47$\pm$0.01 & 19.12$\pm$0.01 & 18.92$\pm$0.01 \\
& G3 & -2.798$\pm$0.005 & 5.631$\pm$0.005 & --- & 22.31$\pm$0.05 & 20.54$\pm$0.01 & 19.95$\pm$0.01 & 19.47$\pm$0.01 & 19.34$\pm$0.03 \\
\hline
J0407-1931 & A &  -0.179$\pm$0.002  &  -1.530$\pm$0.002  &  4.22 & 20.42$\pm$0.08 & 20.34$\pm$0.05 & 20.23$\pm$0.06 & 20.05$\pm$0.04 & 20.00$\pm$0.18 \\
& B &  0.14$\pm$0.01  &  1.04$\pm$0.01   &  1.0 & 21.58$\pm$0.17 & 21.89$\pm$0.05 & 21.44$\pm$0.70 & 21.56$\pm$0.09 & 21.62$\pm$1.08 \\
& G1 &  0.037$\pm$0.003  &  0.488$\pm$0.003  & --- & 20.99$\pm$0.01 & 19.35$\pm$0.00 & 18.93$\pm$0.00 & 18.66$\pm$0.01 & 18.50$\pm$0.03 \\

\hline
J0501-4118 & A &  -1.247$\pm$0.001  &  -1.050$\pm$0.001  &  1.20 & 18.84$\pm$0.01 & 18.79$\pm$0.01 & 18.61$\pm$0.01 & 18.38$\pm$0.01 & 18.47$\pm$0.04 \\
& B &  2.396$\pm$0.001  &  -1.680$\pm$0.001   &  1.0 & 19.10$\pm$0.01 & 19.06$\pm$0.01 & 18.88$\pm$0.01 & 18.68$\pm$0.04 & 18.75$\pm$0.07 \\
& G1 &  1.538$\pm$0.005  &  -0.754$\pm$0.005  & --- & 21.08$\pm$0.02 & 19.52$\pm$0.01 & 18.94$\pm$0.01 & 18.52$\pm$0.01 & 18.48$\pm$0.01 \\

\hline
J0600-4649 & A &  -1.392$\pm$0.002  &  -0.223$\pm$0.002  &  6.40 & 19.36$\pm$0.01 & 19.34$\pm$0.0 & 19.17$\pm$0.06 & 18.90$\pm$0.05 & 19.04$\pm$0.07 \\
& B &  0.948$\pm$0.009  &  0.138$\pm$0.009   &  1.0 & 21.66$\pm$0.01 & 21.58$\pm$0.0 & 21.40$\pm$0.12 & 21.09$\pm$0.08 & 21.18$\pm$0.21 \\
& G1 &  0.443$\pm$0.010  &  0.086$\pm$0.007  & --- & --- & 20.61$\pm$0.02 & 19.54$\pm$0.01 & 19.12$\pm$0.01 & 19.01$\pm$0.01 \\
\hline
J2349-4518 & A &  -1.399$\pm$0.001  &  -0.619$\pm$0.001  &  7.45 & 18.83$\pm$0.01 & 18.62$\pm$0.01 & 18.81$\pm$0.03 & 18.29$\pm$0.05 & --- \\
& B &  0.705$\pm$0.001  &  -0.217$\pm$0.001   &  1.0 & 21.25$\pm$0.01 & 20.90$\pm$0.01 & 21.05$\pm$0.04 & 20.47$\pm$0.06 & --- \\
& G1 &  0.00$\pm$0.03  &  0.00$\pm$0.02  & --- & 22.55$\pm$0.37 & 21.73$\pm$0.10 & 20.62$\pm$0.05 & 20.00$\pm$0.04 & --- \\
\hline

	\end{tabular}
\end{table*}

\clearpage

\twocolumn{\noindent{\textbf{Affiliations}}\\
$^{1}$Institute of Astronomy, University of Cambridge, Madingley Road, Cambridge CB3 0HA, UK\\
$^{2}$Kavli  Institute  for  Cosmology,  University  of  Cambridge,  Madingley Road, Cambridge CB3 0HA, UK\\
$^{3}$Institute of Physics, Laboratoire d'Astrophysique, Ecole Polytechnique  F\'ed\'erale de Lausanne (EPFL), Observatoire de Sauverny, CH-1290 Versoix, Switzerland\\
$^{4}$Departamento de Ciencias Fisicas, Universidad Andres Bello, Fernandez Concha 700, Las Condes, Santiago, Chile\\
$^{5}$Millennium Institute of Astrophysics, Monse\~{n}or Nuncio Sotero Sanz 100, Oficina 104, 7500011 Providencia, Santiago, Chile\\
$^{6}$Physics Department, UC Davis, 1 Shields Ave., Davis, CA 95616\\
$^{7}$Instituto de F\'isica y Astronom\'ia, Universidad de Valpara\'iso, Avda. Gran Breta\~{n}a 1111, Playa Ancha, Valpara\'iso, 2360102, Chile\\
$^{8}$Physics and Astronomy Department University of California Los Angeles, CA, 90095\\
$^{9}$DARK, Niels Bohr Institute, University of Copenhagen, Lyngbyvej 2, DK-2100 Copenhagen, Denmark\\
$^{10}$Fermi National Accelerator Laboratory, P. O. Box 500, Batavia, IL 60510, USA\\
$^{11}$MIT Kavli Institute for Astrophysics and Space Research, Cambridge, MA 02139, USA\\
$^{12}$Kavli Institute for Particle Astrophysics and Cosmology and Department of Physics, Stanford University, Stanford, CA 94305, USA\\
$^{13}$Institute of Cosmology \& Gravitation, University of Portsmouth, Portsmouth, PO1 3FX, UK\\
$^{14}$National Astronomical Observatory of Japan, 2-21-1 Osawa, Mitaka, Tokyo 181-8588, Japan\\
$^{15}$Subaru Telescope, National Astronomical Observatory of Japan, 650 N Aohoku Pl, Hilo, HI 96720, USA\\
$^{16}$Cerro Tololo Inter-American Observatory, National Optical Astronomy Observatory, Casilla 603, La Serena, Chile\\
$^{17}$Instituto de Fisica Teorica UAM/CSIC, Universidad Autonoma de Madrid, 28049 Madrid, Spain\\
$^{18}$CNRS, UMR 7095, Institut d'Astrophysique de Paris, F-75014, Paris, France\\
$^{19}$Sorbonne Universit\'es, UPMC Univ Paris 06, UMR 7095, Institut d'Astrophysique de Paris, F-75014, Paris, France\\
$^{20}$Department of Physics \& Astronomy, University College London, Gower Street, London, WC1E 6BT, UK\\
$^{21}$SLAC National Accelerator Laboratory, Menlo Park, CA 94025, USA\\
$^{22}$Centro de Investigaciones Energ\'eticas, Medioambientales y Tecnol\'ogicas (CIEMAT), Madrid, Spain\\
$^{23}$Laborat\'orio Interinstitucional de e-Astronomia - LIneA, Rua Gal. Jos\'e Cristino 77, Rio de Janeiro, RJ - 20921-400, Brazil\\
$^{24}$Department of Astronomy, University of Illinois at Urbana-Champaign, 1002 W. Green Street, Urbana, IL 61801, USA\\
$^{25}$National Center for Supercomputing Applications, 1205 West Clark St., Urbana, IL 61801, USA\\
$^{26}$Institut de F\'{\i}sica d'Altes Energies (IFAE), The Barcelona Institute of Science and Technology, Campus UAB, 08193 Bellaterra (Barcelona) Spain\\
$^{27}$INAF-Osservatorio Astronomico di Trieste, via G. B. Tiepolo 11, I-34143 Trieste, Italy\\
$^{28}$Institute for Fundamental Physics of the Universe, Via Beirut 2, 34014 Trieste, Italy\\
$^{29}$Observat\'orio Nacional, Rua Gal. Jos\'e Cristino 77, Rio de Janeiro, RJ - 20921-400, Brazil\\
$^{30}$Department of Physics, IIT Hyderabad, Kandi, Telangana 502285, India\\
$^{31}$Department of Astronomy/Steward Observatory, University of Arizona, 933 North Cherry Avenue, Tucson, AZ 85721-0065, USA\\
$^{32}$Jet Propulsion Laboratory, California Institute of Technology, 4800 Oak Grove Dr., Pasadena, CA 91109, USA\\
$^{33}$Kavli Institute for Cosmological Physics, University of Chicago, Chicago, IL 60637, USA\\
$^{34}$Institut d'Estudis Espacials de Catalunya (IEEC), 08034 Barcelona, Spain\\
$^{35}$Institute of Space Sciences (ICE, CSIC),  Campus UAB, Carrer de Can Magrans, s/n,  08193 Barcelona, Spain\\
$^{36}$Department of Astronomy, University of Michigan, Ann Arbor, MI 48109, USA\\
$^{37}$Department of Physics, University of Michigan, Ann Arbor, MI 48109, USA\\
$^{38}$Department of Physics, Stanford University, 382 Via Pueblo Mall, Stanford, CA 94305, USA\\
$^{39}$Center for Cosmology and Astro-Particle Physics, The Ohio State University, Columbus, OH 43210, USA\\
$^{40}$Department of Physics, The Ohio State University, Columbus, OH 43210, USA\\
$^{41}$Center for Astrophysics $\vert$ Harvard \& Smithsonian, 60 Garden Street, Cambridge, MA 02138, USA\\
$^{42}$Lawrence Berkeley National Laboratory, 1 Cyclotron Road, Berkeley, CA 94720, USA\\
$^{43}$Australian Astronomical Optics, Macquarie University, North Ryde, NSW 2113, Australia\\
$^{44}$Lowell Observatory, 1400 Mars Hill Rd, Flagstaff, AZ 86001, USA\\
$^{45}$Departamento de F\'isica Matem\'atica, Instituto de F\'isica, Universidade de S\~ao Paulo, CP 66318, S\~ao Paulo, SP, 05314-970, Brazil\\
$^{46}$Department of Physics and Astronomy, University of Pennsylvania, Philadelphia, PA 19104, USA\\
$^{47}$George P. and Cynthia Woods Mitchell Institute for Fundamental Physics and Astronomy, and Department of Physics and Astronomy, Texas A\&M University, College Station, TX 77843,  USA\\
$^{48}$Instituci\'o Catalana de Recerca i Estudis Avan\c{c}ats, E-08010 Barcelona, Spain\\
$^{49}$Department of Astrophysical Sciences, Princeton University, Peyton Hall, Princeton, NJ 08544, USA\\
$^{50}$School of Physics and Astronomy, University of Southampton,  Southampton, SO17 1BJ, UK\\
$^{51}$Brandeis University, Physics Department, 415 South Street, Waltham MA 02453\\
$^{52}$Computer Science and Mathematics Division, Oak Ridge National Laboratory, Oak Ridge, TN 37831}
\bsp	
\label{lastpage}
\end{document}